\RequirePackage{amsmath}
\documentclass[11pt,paper=a4]{article}
\usepackage{natbib}
\usepackage{makeidx}     
\usepackage{graphicx}    
\usepackage[ansinew]{inputenc}
\usepackage{hyperref}
\usepackage{setspace}
\makeindex             
\catcode`\°=\active\def°{\thinspace}
\DeclareRobustCommand{\Ear}{\raisebox{-0.3em}{\includegraphics[height=1.25em]{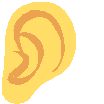}}}
\DeclareRobustCommand{\Arm}{\raisebox{-0.3em}{\includegraphics[height=1.25em]{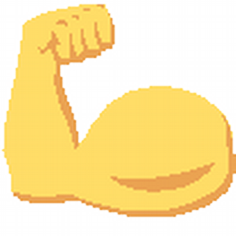}}}
\DeclareRobustCommand{\Eye}{\raisebox{-0.3em}{\includegraphics[height=1.25em]{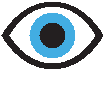}}}
\DeclareRobustCommand{\Person}{\raisebox{-0.3em}{\includegraphics[height=1.25em]{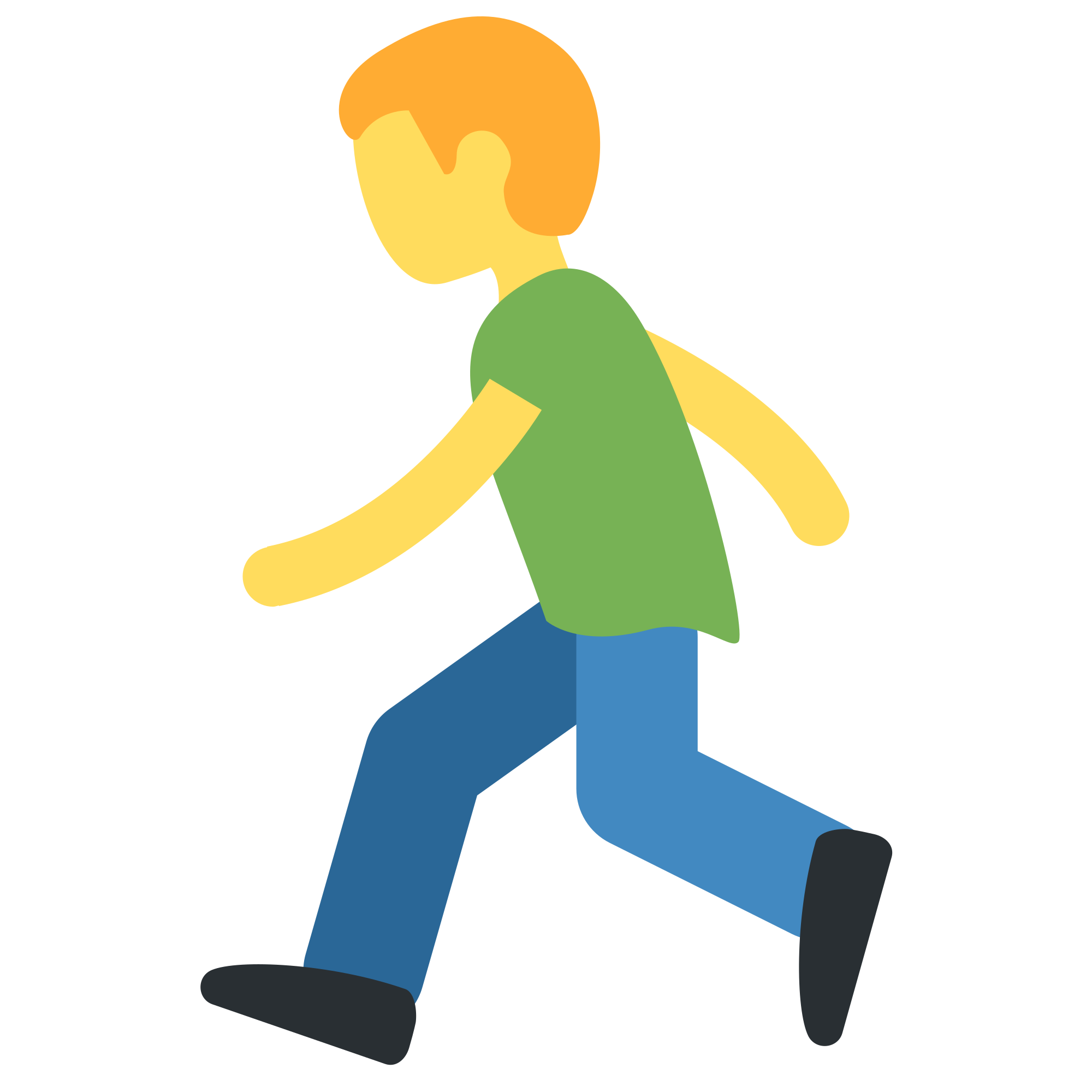}}}

\begin{document}
\title{Formation of three-dimensional auditory space}

\author{Piotr Majdak$^{*}$, Robert Baumgartner\thanks{contributed equally; piotr.majdak@oeaw.ac.at; robert.baumgartner@oeaw.ac.at} , Claudia Jenny}
\date{\small Acoustics Research Institute, Austrian Academy of Sciences, Vienna, Austria} 

\maketitle              
\begin{abstract}
Human listeners need to permanently interact with their three-dimen\-sional (3-D) environment. To this end, they require efficient perceptual mechanisms to form a sufficiently accurate 3-D auditory space. In this chapter, we discuss the formation of the 3-D auditory space from various perspectives. The aim is to show the link between cognition, acoustics, neurophysiology, and psychophysics, when it comes to spatial hearing. First, we present recent cognitive concepts for creating internal models of the complex auditory environment. Second, we describe the acoustic signals available at our ears and discuss the spatial information they convey. Third, we look into neurophysiology, seeking for the neural substrates of the 3-D auditory space. Finally, we elaborate on psychophysical spatial tasks and percepts that are possible just because of the formation of the auditory space.
\end{abstract}
\pagebreak
\section{Introduction}
A loud roar and you turn around. Fight or flight? An archaic situation, typical for our ancestors and animals still living in their original environment. This concept still applies when transferred to the modern human. “Hi!” and you turn around - a good friend has just recognized you on a street. Before you change your walking direction, a car honk lets you look back - you’ve just missed that car crossing your path. 

Such situations make it obvious: in the jungle and on the street, human and non-human animals both need a good understanding of the 3-D world by means of auditory perception. Hearing allows us to create a map of our environment in order to react in a proper way. Towards this goal, the auditory system has to answer the question “what is where?”. 

In this chapter, we review recent advances in understanding how human listeners form and use the auditory space - from the cognitive, acoustic, neurophysiological, and psychophysical perspectives. To this end, in Sec.~\ref{Sec:cognition}, we describe the problem and elaborate on the potential solutions given by researchers from cognitive psychology. As the perceptual outcome depends on the quality with which the acoustic spatial information is conveyed, in Sec.~\ref{Sec:acoustics}, we describe how this information is encoded within binaural signals. The solution to the problem should parallel the neural processing actually happening in our auditory system, and thus, in Sec.~\ref{Sec:neuro}, we briefly describe the neural processing of the acoustic signals while focusing on the extraction of spatial information. Finally, the result of that processing reflects the variety of spatially oriented tasks human listeners can complete. In Sec.~\ref{Sec:Psycho}, we describe various psychophysical spatial tasks demonstrating our abilities to utilize our understanding of the 3-D auditory world.

\section{Cognition: representing the world}\label{Sec:cognition}
From the cognitive perspective, the auditory system not only needs to address the question “what is where?” in order to form a mental representation of our environment because sound is ephemeral, i.e., is happening and short-lived; it is an effect of events, providing information on what happens right now instead of a long-lasting description of objects’ properties. Thus, the information carried by the sound not only needs to be stored for its processing, it also requires consideration of various time scales. On a short time scale, a crack might mean someone stepping on the floor. Many similar cracks, however, would indicate someone opening a door. Thus, the ephemeral property of sound requires the auditory system to address the question “what is happening?”. Answering that question can be only a basis for understanding a static environment. Our world, however, is dynamic and we continuously interact with it. In order to decide which actions have to be done, our auditory system has to provide a basis for the prediction of “what will happen next?”.

\subsection{The ill-posed problem}\label{SubSec:Ill-posed}
\begin{figure}
	\centerline{\includegraphics[width=1.0\textwidth]{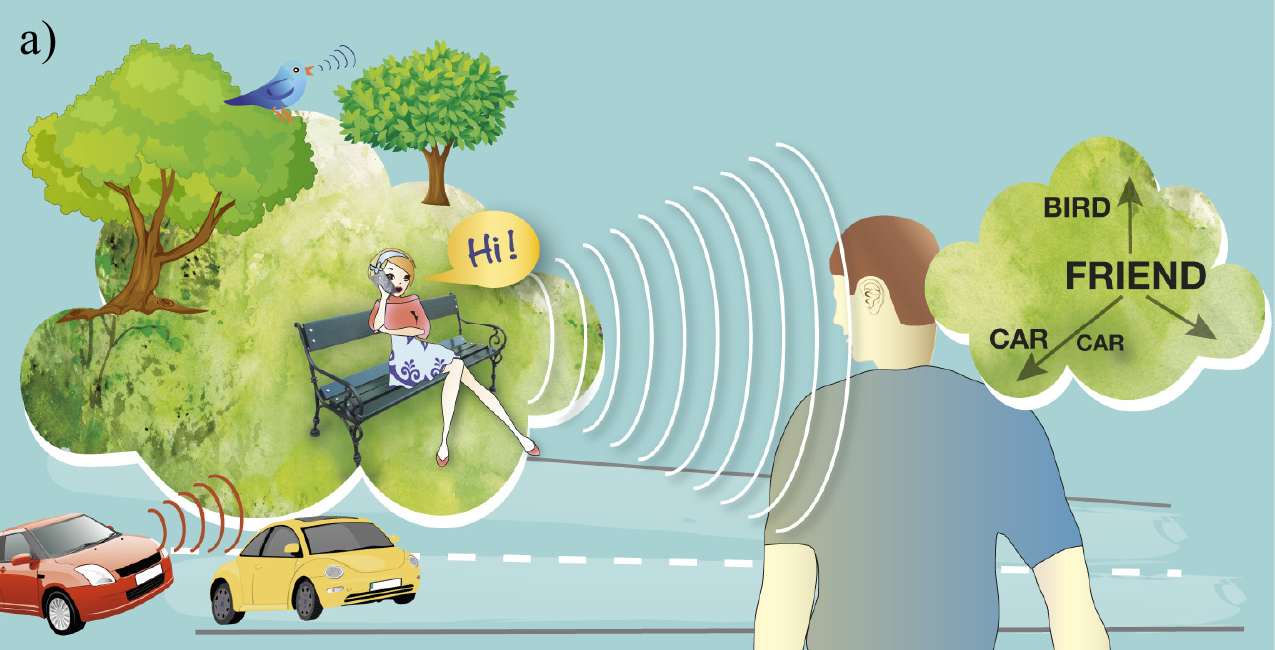}}
	\centerline{\includegraphics[width=1.0\textwidth]{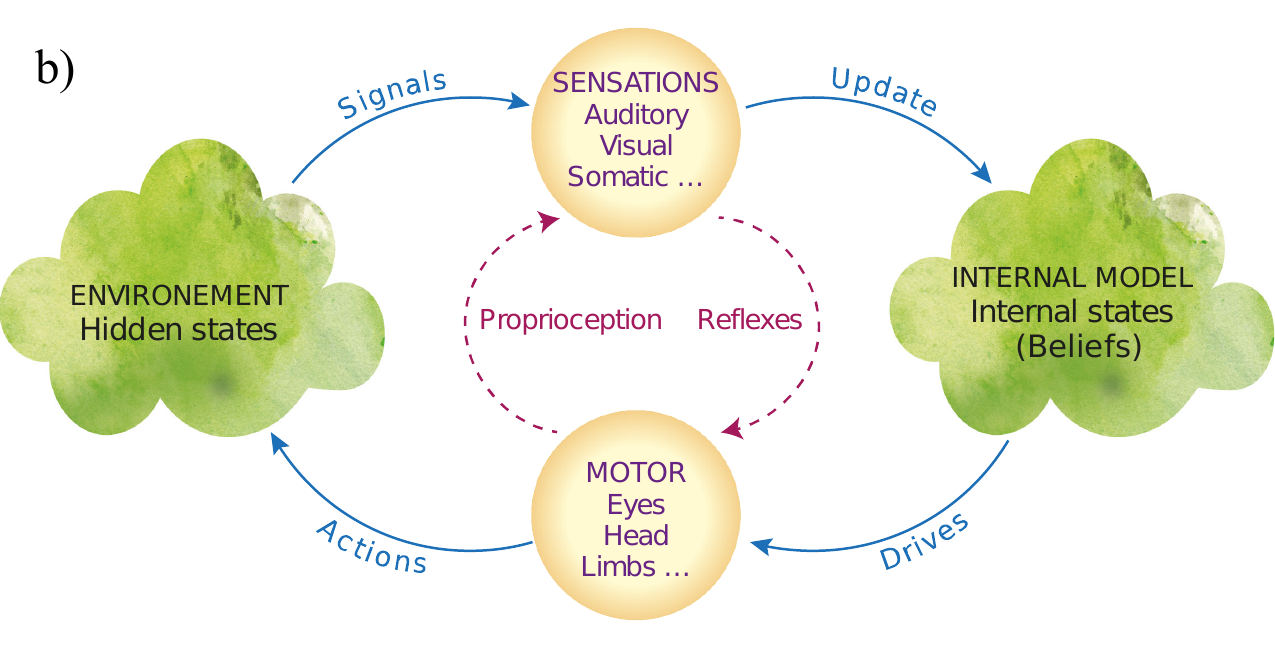}}
	\caption{\textbf{a)} Perception as a mental representation of the environment based on sensation. The objects (left) producing sounds that are perceived by a listener (center) are represented in an internal model (right).\textbf{ b)} Active inference. Sensations (top) of signals from the environment (left) are used to update the internal model beliefs (right) on the hidden states of the environment. The model predictions drive actions (bottom) which allow to interact with the environment.}
	\label{Fig:cognition}
\end{figure}

Auditory scenes consist of objects producing sounds (see Fig. \ref{Fig:cognition}a). Perception, as a process of transforming sensory information to higher levels of representation, needs to mentally represent these objects and their properties. Here, an “auditory object” can be thought of as a perceptual construct linking a sound with a corresponding source \citep{griffiths_what_2004}. Sitting in a park, hearing a honk, a word, and a chirp would let us identify auditory objects of a car, a human, and a bird. Auditory objects can be assigned to a spatial position, among other non-spatial properties. Now, when the human starts to speak, and the bird starts to sing, these two objects become sources of acoustic streams. Their mixture arrives at our ears and the job of the auditory system is to separate them into two auditory streams, which can be defined as a series of coherent events that can be grouped and attributed to a single auditory object. Hence, the formation of the auditory space depends on the capabilities of the listener to form auditory objects and estimate their spatial properties. 

As most of the sound sources are usually located outside of our body, the formation of the auditory space can be seen as a perceptual task aiming at the reconstruction of the external (or distal) state of affairs \citep{epstein_perception_1995}. Unfortunately, this is an ill-posed problem emerging from the fact that not only does the head-centered binaural signal need to be transformed to a world-centered representation, but also that the information available at our ears is insufficient to exactly reconstruct a unique state. This is because a given binaural signal arriving at our ear drums can originate from an infinite number of sound-producing events, all of which having produced exactly that signal. This ill-posed problem results, as a consequence, in ambiguity in creating auditory objects.

A famous example showing the difficulty of mapping acoustics to a sound-producing event is the estimation of a 2-D surface given a sound wave, posed as “can we hear the shape of a drum?”. Indeed, it has been formally shown that for many shapes, one cannot completely differentiate the shape of the drum because an exact solution of the plane geometry from the waveform is impossible \citep{gordon_one_1992}. However, additional or even redundant information can help and, fortunately, our acoustic environment is full of redundancy. Recent developments in mathematical tools show the variety of tricks the auditory system may potentially use when it comes to utilizing acoustic redundancy and thus better solving inverse problems. For example, with a known but monaural only signal, the shape of any convex room can be estimated from the delays between the early reflections in the room impulse response \citep{moore_room_2013}. Given four acoustic sensors, the room shape can be estimated by relying on the delay of just the first reflections \citep{dokmanic_acoustic_2013}, showing the power of better solutions given redundancy (here, multiple sensors). Thus, it is not surprising that binaural hearing, by providing interaural information as compared to monaural hearing, allows us, to better retrieve the spatial properties of the environment, better orient ourselves within the environment, and even improve speech perception \citep[][and also Clapp and Seeber, this volume]{bronkhorst_cocktail-party_2015}. 

Despite the redundancy in the binaural signal, solving the ill-posed problem requires certain assumptions in order to reduce the infinite number of potential solutions \citep[e.g.,][]{friston_embodied_2012}. Generally, these assumptions are driven by the goal of an efficient interaction with the environment sampled by the sensors. At the end of the cognitive process, the solution needs to provide a basis for decisions that trigger the execution of appropriate actions. If the vast amount of sensory information was processed from scratch each time an action was required, most of the actions would happen too late. Faster processing can be achieved with predictive coding by introducing an internal model \citep{francis_internal_1976} that predicts the external state of affairs and is continuously adapted based on incoming sensory data \citep[for more details on predictive coding, see][]{aitchison_or_2017}. Such models have been introduced in motor-control theory and robotics to describe reaching movements, to plan movement trajectories, and to model imagery \citep[for review and discussion, see][]{grush_emulation_2004}. In cognition, the term "perceptual inference" has been coined \citep[e.g.,][]{hinton_generative_1997}. 

In predictive coding, the more realistic the model predictions, the more efficiently actions can be performed. In other words, the objective of the internal model is to minimize surprise, which then yields minimal corrections to be applied to the performed actions. In that sense, the process of cognition can be considered as forming a generator creating hypotheses and then testing them against the pre-processed sensory information \citep{gregory_perceptions_1980}. The free-energy principle has been proposed to explain how the cognitive system can efficiently create a model predicting the environment while restricting itself to a limited number of states \citep{friston_free_2006}. Free energy, a concept with a long tradition in thermodynamics \citep{helmholtz_sensations_1954}, is the difference between the internal energy of a system and the energy required to describe the actual states of that system. In the free-energy concept, the lower is the free energy, the more efficiently is the system described. In terms of cognition, a cognitive model creates plausible predictions that minimize surprise given the states described by the present sensory information and the internal states describing the model’s beliefs about the environment (hidden states). The free energy acts as a prediction error and is minimized by choosing that most plausible prediction which most efficiently drives the motor system. The internal states of the model are then updated based on the new sensory information about the hidden states of the environment. This process could be termed "active inference" \citep[for review, see][]{friston_active_2016}. 

\subsection{Rules and limitations}

In active inference, a model‘s beliefs represent rules describing plausible environments. They can be learned throughout the development \citep{bhatt_how_2011} and limit the potential solutions to the plausible ones only. This application of limitations has several implications. First, they sometimes fail, yielding unrealistic representations. Illusions, i.e., distortions of the perceived physical reality, are great examples of the consequences of plausible but wrong assumptions while solving the ill-posed problem \citep[e.g.,][]{carbon_understanding_2014}. Understanding their origin can help uncover the underlying processes in auditory perception. Second, these limitations allow us to reduce the vast amount of sensory information to a smaller number of informational units along the ascending pathways of processing. In our case, a small and discrete number of auditory objects with a finite number of properties is created from a continuous binaural signal. In the process, the frame of reference undergoes a transformation from the head-centered binaural information to world-centered information about our environment \citep{schechtman_spatial_2012}. 

Interestingly, this whole process can be seen as a nonlinear extension of compressed sensing, a signal processing technique for efficiently acquiring and reconstructing a signal by finding solutions to underdetermined linear systems \citep{donoho_for_2006}. In compressed sensing, the constraint of sparsity is chosen in order to find a solution to the underdetermined system. Compressed sensing is widely used in signal processing, but it requires a \textit{linear relation} between the observation and solution. In active inference, this process is described by means of variational \textit{Bayes statistics}, which aims to provide an analytical approximation to the posterior probability of the unobserved variables in order to obtain statistical inference about these variables. 

At the end, it is all about reducing the amount of sensory information. A widely accepted concept describing the reduction of auditory information to discrete informational units is auditory scene analysis \citep[ASA, see van de Par et al., in this volume and ][]{bregman_auditory_1990}. ASA assumes that our auditory system partitions the acoustic signal into auditory streams that each refer to an auditory object. While good separability between fore- and background streams was originally believed to constitute the goal \citep{bregman_auditory_1990}, more recent models consider their predictability as the main motivation for grouping \citep{winkler_modeling_2009}. Grouping mechanisms seem to rely on auditory features like onset, pitch, spectrum, and interaural disparity, and can act simultaneously and sequentially. Simultaneous grouping assumes that features are integrated to a foreground property, like harmonics coming from the same instrument are integrated to a single pitch, and features deviating from the expected and learned patterns are segregated and form a background. Sequential grouping integrates and segregates auditory objects and streams, depending on their temporal properties. It can even override the result of simultaneous grouping: For example, sounds simultaneously presented with a different interaural disparity can be grouped into a single auditory stream, but the same sounds embedded in an acoustic stream presented from the spatial location corresponding to one of the sounds can create two auditory objects appearing at the distinct spatial locations \citep{best_binaural_2007}.

Bregman’s ASA further involves the concept (“old-plus-new”) of competitive processes determining the auditory stream reaching awareness. These processes have been originally derived as obeying the laws of the Gestalt theory, which provides a description of our ability to acquire plausible perceptions from the sensory input \citep{koffka_principles_1935}. The main assumption of the Gestalt theory is that our perception is influenced by perceptual units derived from the signal according to the laws of proximity, similarity, closure, symmetry, common fate, continuity, good Gestalt, and past experience. Even though the Gestalt theory has difficulties in providing insights into the neural processes leading to perception \citep{schultz_history_2015}, the laws of the Gestalt theory helped in constraining the ambiguity resulting from the ill-posed problem of perception. Recent findings show that these neural processes are based on heuristics acquired through learning and experience \citep{shinn-cunningham_object-based_2008}. Further, these processes act “top-down” and can be modulated by other modalities like vision \citep[yielding ventriloquism or self-motion;][]{kondo_effects_2012} or auditory attention \citep{hill_auditory_2010}. 

Depending on the relevance of top-down modulations, neural processes may be distinguished as being reflexive or reflective. Reflexive processes result in very fast reactions (with latencies below 100 ms), which can usually not be suppressed \citep{curtis_success_2003}. They involve startle reflex or orienting reflex \citep{sokolov_orienting_2001}, and do not require, but can be modulated by attention. They can be used to trigger movements toward auditory sources or intensify the processing of cues that signal approaching objects \citep{baumgartner_asymmetries_2017}. Thus, they are vital in protecting humans from hazardous events. In contrast, reflective processes have longer latencies, require top-down attention, a controlled bias in the preference for and processing of the information streams \citep[for review, see][]{knudsen_fundamental_2007}. Attention is usually thought to be a single unidirectional process representing task-specific goals and expectations \citep[top-down,][]{awh_top-down_2012}, however, it can also be modulated by various components of a stimulus \citep[bottom-up,][]{arnal_human_2015}, i.e., salient components. Reflective or attentional processes require working memory to develop and test hypotheses based on the salience in a stimulus \citep{carlile_selective_2015}. A similar distinction between reflexive and reflective processes has also been proposed for speech category learning \citep{chandrasekaran_toward_2014} and has been used to describe social behavior \citep{strack_reflective_2004}. 

While the reflexive processes in auditory processing have been widely investigated, the actual reflective processes behind the ASA are not completely understood yet. Thus, it is not surprising that there is much debate on the mechanisms underlying ASA, which has been widely discussed from different perspectives \citep[e.g.,][]{bizley_what_2013,szabo_computational_2016,micheyl_role_2007,nelken_auditory_2014,snyder_recent_2017}.

In summary, the free-energy concept provides a solid statistical framework for deriving the external state of affairs, and ASA provides a valid conceptual framework for the cognitive processing of auditory information (for more details, see van de Par, this volume). The particular result in terms of a realistic representation of our world depends on the quality with which the spatial information about the auditory objects is conveyed by the binaural signal. Thus, in the following section, we describe the acoustic spatial information encoded in binaural signals.

\section{Acoustics: formation of binaural signals}\label{Sec:acoustics}
The sounds arriving at our ear drums are acoustically filtered versions of the sound actually produced in our environment. The filtering is a result of the interaction of the sound field with the reverberant space and our body parts such as the head, torso, pinna, and ear canal. 

\begin{figure}
	\centerline{\includegraphics[width=1.0\textwidth]{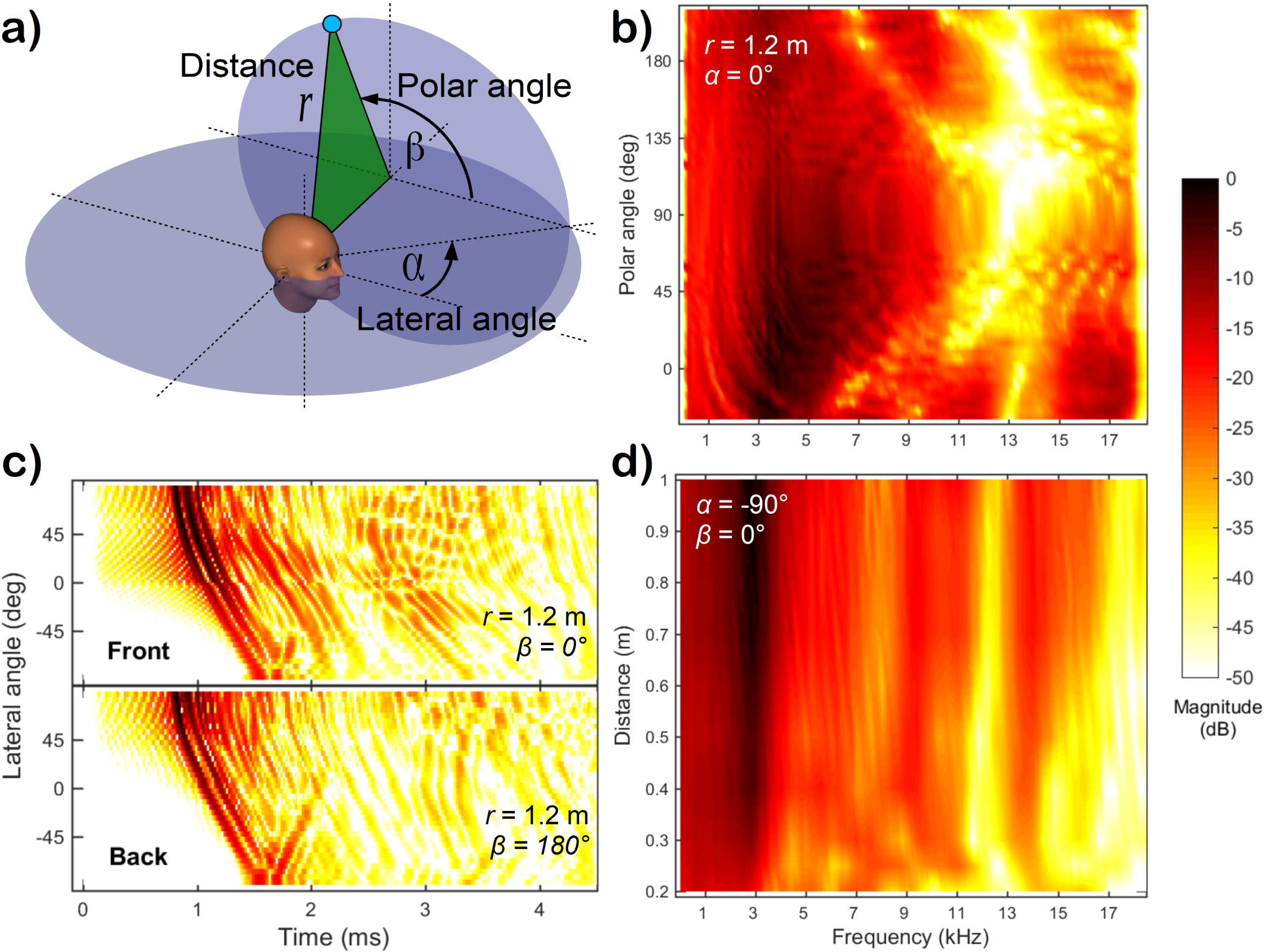}}
	\caption{\textbf{a)} Interaural-polar coordinate system with the lateral angle $\alpha$, the polar angle $\beta$, and distance $r$; \textbf{b)} magnitude spectra of far-field HRTFs along the median plane as a function of the polar angle; \textbf{c)} energy-time curves (ETCs) of far-field HRIRs of a left ear along the lateral angle to the front (top) and back (bottom) of the horizontal plane; \textbf{d)} magnitude spectra of near-field HRTFs of a left ear for the direction ($\alpha,\beta$): ($-90^\circ,0^\circ$) (the most-right direction at the horizontal plane) as a function of distance. Color bar: magnitude in dB.}
	\label{Fig:monaural}
\end{figure}

In acoustics, sound fields are commonly described in Cartesian or spherical coordinates. In psychoacoustics, however, the interaural-polar coordinate system, Fig. \ref{Fig:monaural}a, better corresponds with the human perception. This system is created by rotating the poles of the spherical system by 90$^\circ$, such that the two poles are aligned with the interaural axis connecting the two ears. In that system, the lateral angle $\alpha$ describes the lateral position of a source, ranges from -90$^\circ$ to +90$^\circ$, selects the sagittal plane (i.e., all parallel planes of the median plane), and for directions in the horizontal frontal half-plane, corresponds to the azimuth angle. The polar angle $\beta$ describes the position of the source along the sagittal plane and ranges from -90$^\circ$ (bottom) via 0$^\circ$ (eye-level, front), 90$^\circ$ (top), and 180$^\circ$ (eye-level, back) to 270$^\circ$ (bottom again). Together with the distance $r$, we use lateral and polar angles throughout this chapter to describe the spatial position of sound sources.

\subsection{Monaural features}
For a sound coming from a particular direction, its filtering can be captured by the binaural pair of head-related transfer functions (HRTFs). While the filtering of the sound happening in the ear canal does not depend on the incidence angle of the sound, the filtering by the head, torso, and pinnae does, creating direction-dependent changes of the received sound. The directional information provided by individual HRTFs is referred to as the monaural spectral features. Their direction-dependent changes are especially apparent for a sound moving along the median plane of the listener (see Fig.~\ref{Fig:monaural}b). The change in the elevation and front/back dimension can be directly linked with the spectral changes, that is, changing notches and peaks as a consequence of cancellation and amplification, respectively, caused by various body parts. The reflections of the torso create spatial frequency modulations up to 3 kHz \citep{algazi_elevation_2001}. The head shadows frequencies above 1 kHz; however, above 6 to 8 kHz, the effect of the pinna is most prominent \citep{blauert_spatial_1997}. The reflections at the pinna create peaks and notches at frequencies above 4\,kHz. For example, the directionality of the pinna towards the front causes an attenuation of high frequencies for sounds coming from behind the listener. This manifests in spectral peaks above 8 kHz for the frontal sound positions. Further, an increase in sound elevation changes the varying delay between the direct sound and its reflection along the pinna concha. This manifests in an upward shift of the spectral notches usually found between 6 and 10 kHz. A further contribution is that of the head. 

In general, the pinna is the reason for a large variation in the HRTFs among listeners, as the pinna’s geometry varies among the healthy human population \citep{algazi_elevation_2001}. While HRTFs are similar across listeners at frequencies up to 6 kHz, differences as large as 20 dB have been found at higher frequencies \citep{moller_head-related_1995}. Listener-specific HRTFs can be acquired by applying system identification approaches on acoustical measurements \citep[for review, see][]{majdak_multiple_2007}. The acoustic measurement including all positions in 3-D space is a resource-demanding procedure and takes tens of minutes, even when sophisticated measurement methods are applied. HRTFs can also be calculated based on a geometric representation of the listener \citep{kreuzer_fast_2009}; however, the demands on geometric accuracy and computational power are high \citep{ziegelwanger_numerical_2015}. Recent developments in the acquisition of the 3-D geometry from photographs by means of photogrammetric reconstruction \citep{reichinger_evaluation_2013} and numeric algorithms \citep{ziegelwanger_priori_2016} seem promising in easing the acquisition of listener-specific HRTFs in the future. While (for research purposes) HRTFs have been measured for a long time, their exchangeability was limited because of missing standards for their representation. Recently, the spatially oriented format for acoustics (SOFA) was created \citep{majdak_spatially_2013} as a standard, easing their exchangibility, and promoting their usage in consumer applications.

HRTFs can also be analyzed in the time domain by applying the inverse Fourier transformation on each HRTF yielding head-related impulse responses (HRIRs, see Fig.~\ref{Fig:monaural}c). HRIRs usually decay within the first 4~ms and show the direction-dependent delay between the sound source and the ear. The temporal position of the first onset in an HRIR can be considered as the broadband time-of-arrival (TOA), which, based on the approximation of the head as a sphere, for a listener, can be modeled as a spatially-continuous 3-D representation of the broadband delay requiring only few parameters \citep{ziegelwanger_modeling_2014}. Even though HRTFs show a nonlinear spectral phase depending on the sound direction, the HRTF phase spectrum can be represented by a combination of the minimum phase derived from the HRTF amplitude and the linear phase corresponding to the TOA \citep{kulkarni_sensitvity_1999}.

HRTFs also vary with distance, especially in the near field (see Fig.~\ref{Fig:monaural}d). This is due to the contribution of the head shadow and changes of the pinna-reflection paths  \citep{brungart_auditory_1999-1}. The nearer the source, the less diffraction around the head occurs at lower frequencies and the less intense are the reflection patterns of the pinna. Low-frequency attenuation of up to 20 dB is a prominent spatial feature encoding distance for near sounds.  

Note that the sound source itself might be a source of spatial cues because the HRTF filtering is commutative, i.e., the auditory system has no chance to distinguish whether the evaluated spectrum originates from the sound source or is an effect of filtering by an HRTF. For example, when listening to a frequency sweep of notched noise, the uprising spectral notch may be the reason for the illusion of an elevating source \citep{blauert_spatial_1997}. Alternately, sounds with spectral ripples may overlap with the monaural spectral features of HRTFs and interfere with the derivation of directional information from the binaural signal \citep{macpherson_vertical-plane_2003}. 

\subsection{Interaural cues}
Having two ears allows us to sense the sound field at two different spatial positions. Thus, we have access not only to monaural features, but also to the combination of the left- and right-ear features, the so-called interaural cues. The most prominent examples of the interaural cues are the broadband interaural time and level differences (ITDs, ILDs). Their importance for sound localization was recognized very early \citep{lord_rayleigh_or_strutt_our_1876}. Later, the general dissimilarity of the signals between the two ears, expressed as binaural incoherence and spectral ILDs, have been found to be important \citep{blauert_spatial_1997}.

\begin{figure}[t]
	\centerline{\includegraphics[width=1.0\textwidth]{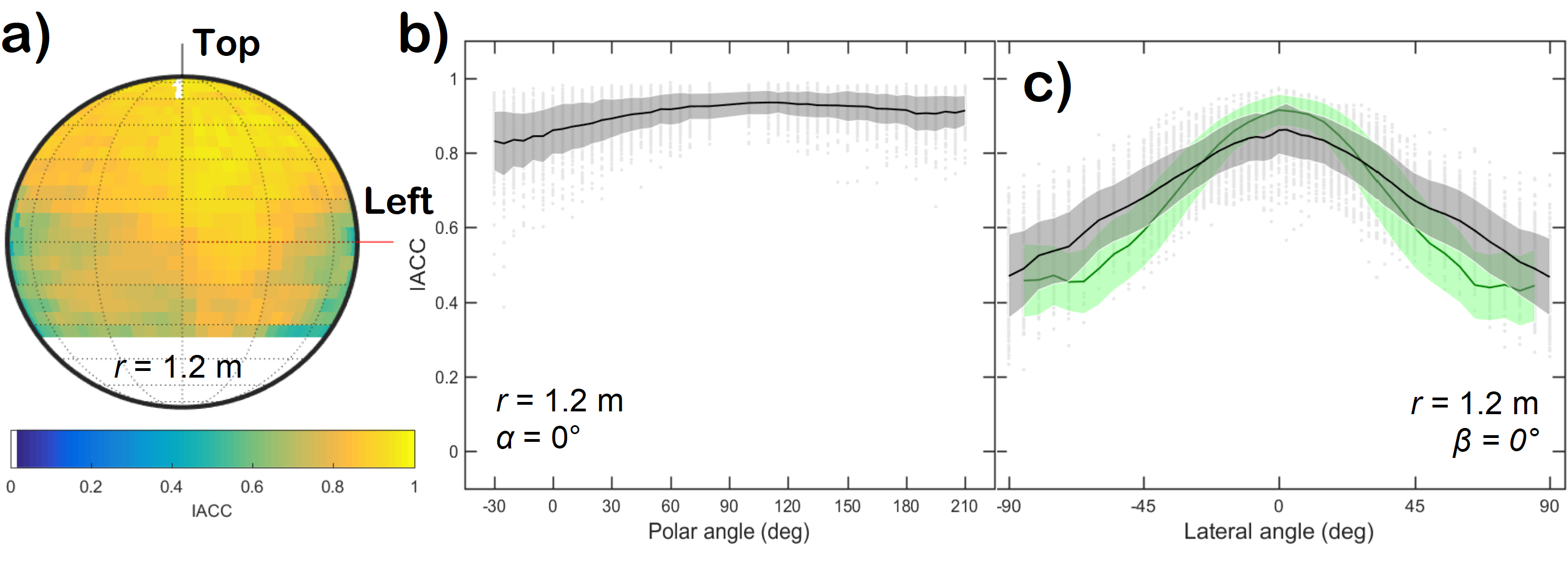}}
	\caption{\textbf{a)} Interaural cross-correlation coefficients (IACCs) of a single listener's HRTFs for various directions when looking at the listener from the front. \textbf{b)} IACCs along the median plane calculated for a listener population. \textbf{c)} IACCs along the horizontal front (gray) and rear (green) half-planes calculated for a listener population. The population consisted of 97 listeners from the ARI database \citep{majdak_3-d_2010}. The dots show the individual IACCs, the line and gray area shows the total average and $\pm 1$ standard deviation, respectively, across the population.}
	\label{Fig:IACC}
\end{figure}

\subsubsection{Interaural signal similarity}
The similarity between the signals of the listener’s two ears has been described by many terms: interaural coherence, interaural cross-correlation, binaural incoherence, interaural decorrelation, or even the binaural quality index. And all of them consider the temporal similarity between the left- and right-ear signals, thus intrinsically describe spectral features. The interaural cross-correlation function $r_{LR}$ of the two signals $x_L$ and $x_R$ is a function of the interaural lag $\tau$ \citep{goupell_interaural_2006}:
\begin{equation}\label{eq:crosscorr}
r_{LR}(\tau)=\frac{\int x_L(t) x_R(t+\tau)\;dt} {\sqrt{\int
		x_L^2(t_1)\;dt_1 \int
		x_R^2(t_2)\;dt_2}},
\end{equation}
\noindent This function can vary between -1 and 1 and typically has a single peak. The lag of that peak corresponds to the broadband ITD and is mostly determined by the lateral position of a sound source. The height of that peak is usually known as the interaural cross-correlation coefficient (IACC) and demonstrates the best interaural similarity of the binaural signal (see Fig.~\ref{Fig:IACC}a). Note that IACCs are naturally below 1. Typically, free-field IACCs, even in the median plane, are around 0.9 (see Fig.~\ref{Fig:IACC}b), because the HRTFs are not identical as a consequence of non-identical ears. In the horizontal plane, the IACC decreases with increasing lateral angle of the sound source, with typical IACCs around 0.4 for the most-lateral directions (see Fig.~\ref{Fig:IACC}c). Note that this is a broadband consideration of the IACC and the interaural dissimilarities may be different and contribute differently across frequencies. 

\subsubsection{ITD}
\begin{figure}
	\centerline{\includegraphics[width=1.0\textwidth]{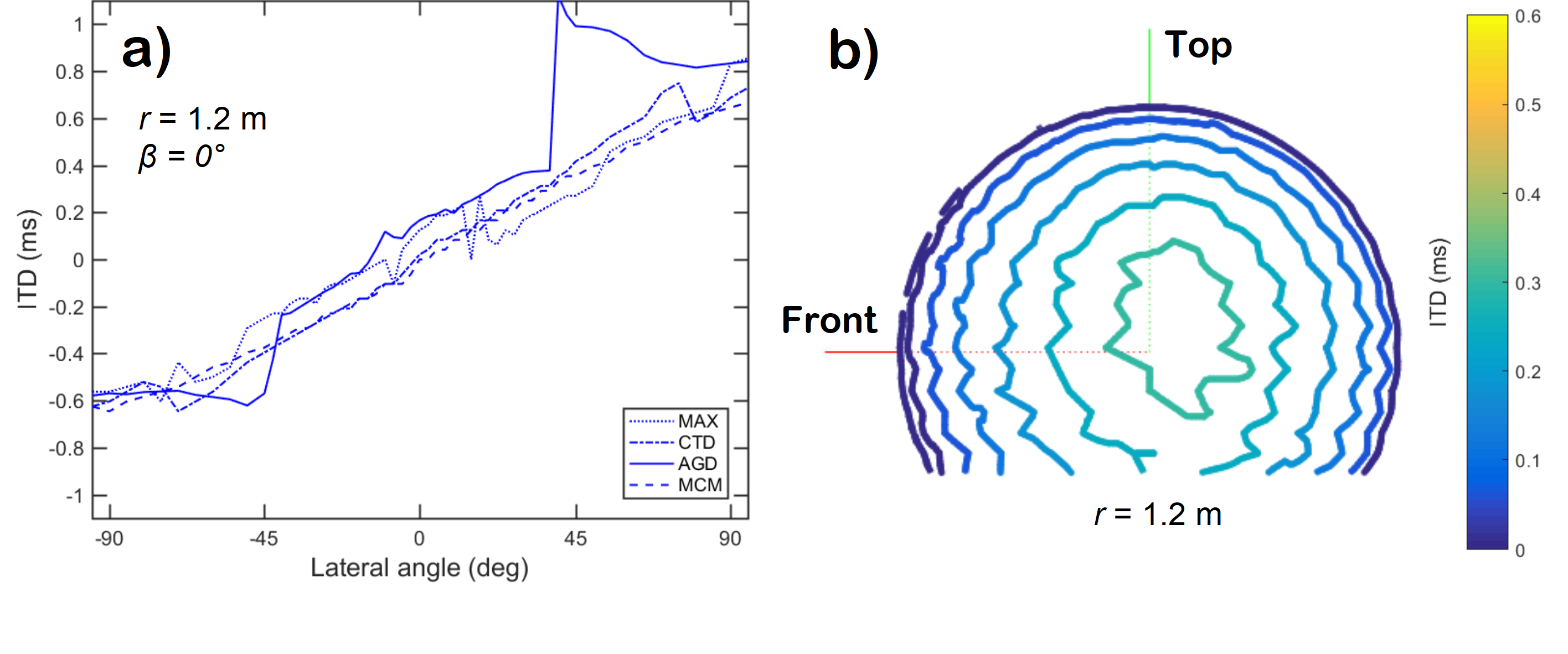}}
	\caption{\textbf{a)} ITDs for frontal directions at the horizontal plane, estimated by various ITD extraction methods (see text). \textbf{b)} Iso-ITD contours; view at the left ear along the interaural axis. Color bar: ITD in ms.}
	\label{Fig:ITD}
\end{figure}

When talking about the ITD, people usually refer to the broadband ITD. However, given the limited interaural coherence and the attendant frequency dependence, a broadband ITD can only be an approximation of delays appearing between the two ears. Consequently, various methods have been proposed for the calculation of the ITD, based on HRTFs represented in the time or frequency domains. Figure \ref{Fig:ITD}a shows the ITDs as a function of the lateral angle obtained from an HRTF set by using various methods. In the time domain, methods either focused on the ITD between the first onsets \citep[MAX in Fig. \ref{Fig:ITD}a, see][]{andreopoulou_identification_2017} and centroids (CTD), or the lag of the IC peak of HRIRs compared to their minimum-phase versions (MCM). In the frequency domain, the ITD can be calculated from the spectral average of the interaural group delay (AGD). While the correct estimation method clearly depends on the application, recently, ITDs between the -30-dB onsets (relative to the peak) in low-pass filtered HRIRs showed the best congruence with psychophysical tasks when tested with broadband sounds in humans \citep{andreopoulou_identification_2017}. From the geometrical perspective, there is a long history of various ITD models based on representations of the head as a circle, sphere, ellipsoid, and polynomial function \citep{xie_head-related_2013}. 

Generally, ITDs are frequency dependent and theoretical considerations show that low-frequency ITDs are 50\% larger than those at higher frequencies \citep{kuhn_model_1977} with a transition around 1.5 kHz. The maximal ITD depends on the listener’s head diameter and the calculation method and has a population average of around 850 $\mu$s \citep{algazi_elevation_2001}. ITDs in that range imply that sounds with frequencies below 1.2 kHz undergo an interaural phase shift of less than $180^\circ$ when traveling from one ear to the other, and the interaural phase difference can unambiguously encode the source direction. At higher frequencies, sounds with wavelengths smaller than the head diameter provide interaural phase differences larger than $180^\circ$ yielding ambiguous ITDs. Hence, for stationary tones, ITDs in higher frequencies do provide unique information about the sound’s lateral direction. In the case of amplitude-modulated and multi-tone sounds, the timing of the envelopes may also be informative even at higher frequencies, yielding envelope ITDs as a useful acoustic feature \citep{henning_detectability_1974}.

But even broadband ITDs do not provide much information about the sound’s spatial position beyond its lateral direction. Fig. \ref{Fig:ITD}b shows contours of iso-ITDs derived from an HRTF set of a listener. The contours approximate sagittal planes, demonstrating that ITDs are not able to encode the sound’s distance and elevation, nor enable discrimination between front and back. This finding is not new, as Lord Rayleigh mentioned already in 1876: \textit{“The possibility of distinguishing a voice in front from a voice behind would thus appear to depend on the compound character of the sound in the way that it is not easy to understand, and for which the second ear would be of no advantage” }\citep{lord_rayleigh_or_strutt_our_1876}. Nowadays, the spatial ambiguity based on the ITD is called the "cone of confusion", or "torus of confusion" when distance is involved \citep{shinn-cunningham_tori_2000}.

\subsubsection{ILD}
\begin{figure}
	\centerline{\includegraphics[width=1.0\textwidth]{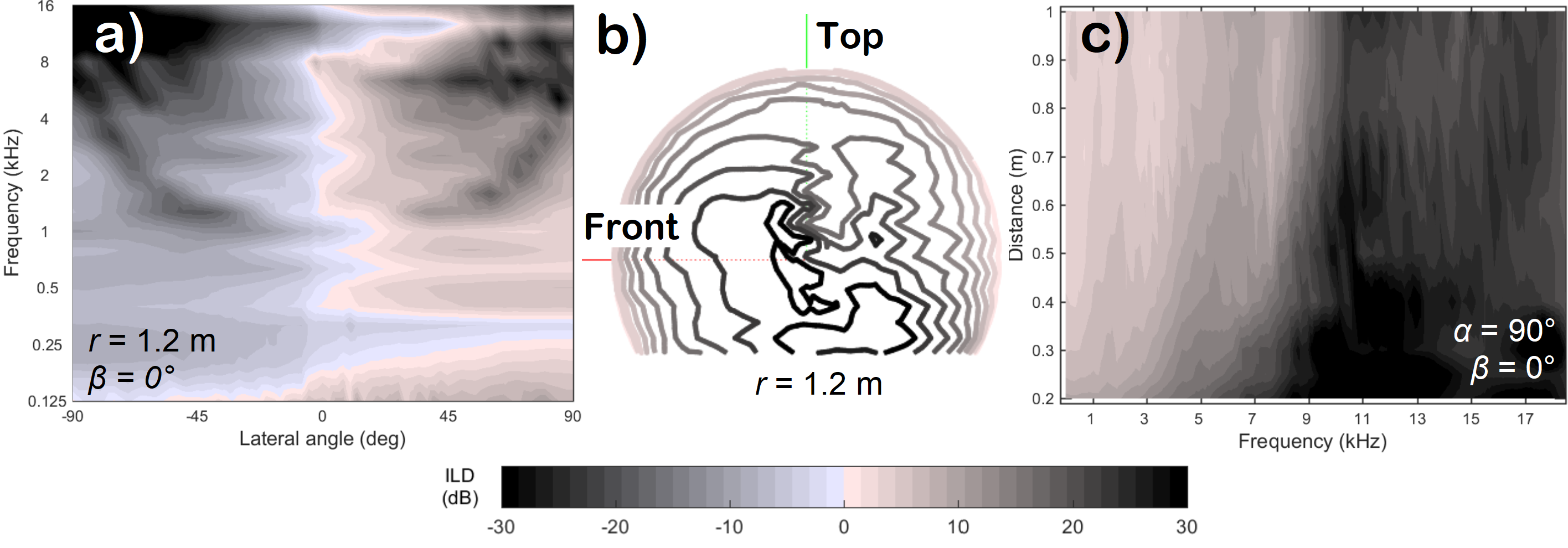}}
	\caption{\textbf{a)} Frequency-dependent ILDs for sounds along the frontal horizontal half-plane. \textbf{b)} Iso-ILD contours; view at the left ear along the interaural axis. \textbf{c)} Frequency-dependent ILDs for the most-left direction in the horizontal plane as a function of distance. Color bar: ILD in dB. Frequency dependence shown by filtering each HRTF by a typical Gammatone filter.} 
	\label{Fig:ILD}
\end{figure}

ILDs arise because of two effects. First, the head is an obstacle, creating a shadow for the contralateral ear, and thus ILDs. Because of that, ILDs increase with both frequency and lateral angle (see Fig.~\ref{Fig:ILD}a), defined by the relation between the sound's wavelength and the listener's head size. While low-frequency ILDs span a range of $\pm$5\,dB and increase smoothly for more lateral sounds, high-frequency ILDs exhibit a span of $\pm$20\,dB, with a more complex relation to the lateral angle. Second, for near-field sounds, the sound intensity decreases rapidly with the distance to the source, creating an ILD even at frequencies for which the head is acoustically transparent, (see Fig.~\ref{Fig:ILD}c). Such low-frequency near-field ILDs become significant for distances below 0.5\,m \citep{brungart_auditory_1999-1} and can reach even beyond 20\,dB.

Similar to the front-back ambiguity of the ITD, ILDs do not vary consistently with the polar angle (see Fig.~\ref{Fig:ILD}c). Thus, ILDs do not encode source directions along the sagittal planes well either, further contributing to the cone of confusion based on interaural features.

\subsection{Reverberation}
Our considerations so far addressed the binaural signal without taking its origin into account. A single sound source presented in the free field, that is, just direct path between the sound and listener without any reflections, is the simplest case. 

However, most realistic binaural signals involve listening in reverberant spaces like rooms. When we consider the simple case of a source and its reflection from a wall, the binaural signal will contain the direct signal overlapped with filtered versions of itself. The filtering consists of a broadband delay (because of the longer propagation path) and spectral changes (because of the frequency-dependent absorption of the sound on the wall). While the addition of the reflection can be considered an echo, in the frequency domain, it yields a comb filter, i.e., a spectral ripple with the spectral density depending on the delay between the direct and reflected sound.

Note that in realistic situations, upon thousands reflections are created by just a single source. Beginning with the clearly distinguishable early reflections, their temporal density increases such that after some time they become a diffuse field, i.e., a sound field with a statistically constant directional distribution. In addition to specular reflections, diffraction and diffusion contribute to the complexity of the reverberant sound field. 

The sound reflections from walls and other surfaces temporally overlapping with the direct sound, create instantaneous ITD fluctuations. As a consequence, the interaural similarity in the binaural signal and thus the interaural coherence decreases. \citep{hartmann_binaural_2005}. The high-frequency envelopes seem to be less susceptible to the reduction of the interaural coherence caused by reverberation as compared to the low-frequency phase differences \citep{ruggles_why_2012}.

Acoustically, the effect of reverberation can be described by the binaural room impulse response (BRIR), which is basically the HRIR measured in a room of interest. While a binaural pair of HRIRs are given for the relative relation between the source and listener’s positions, BRIRs further depend on the absolute positions of the source and listener in the room. While for HRIRs, the source position change is equivalent to the orientation change of the listener, in BRIRs, these parameters cannot be exchanged. Hence, a BRIR is a function of at least the source position, listener position, listener orientation, room acoustics, and maybe even source orientation. 

The position-dependent effect of reverberation implies that it produces different signals at the two ears. Accordingly, the IACC decreases, for frequencies above 500 Hz, even approaching zero depending on the position of the source and listener in a room \citep{hartmann_binaural_2005}. Sounds from various directions following the direct sound cause a fluctuation of interaural cues over the time course of the BRIRs, mostly manifesting as a time-dependent IACC. Hence, IACCs calculated over various ranges of time (and frequencies) are widely used in room acoustics \citep{mason_frequency_2005}. 

\subsection{Dynamic acoustic situations}
Even though a sound is an ongoing temporal fluctuation of pressure, its spatial properties do not change unless the spatial configuration between the listener, source, and space changes. In spatial hearing, this situation is considered as a static one. A widely investigated case is listening to a static sound source without any head movement. In this situation, the HRTFs (or BRIRs in a room) do not change. When the source or listener changes the spatial position and/or orientation, the listening situation is called “dynamic”. As soon as the listener moves (the body or just the head), the HRTFs (or BRIRs) change, creating a systematic temporal change in spatial cues. 

In order to generally describe spatial changes in listening situations, six degrees of freedom need to be considered for each object. For example, the listener’s head can be rotated along the horizontal plane (to the left and right, changing the yaw), along the median plane (up and down, changing the pitch) and it can be tilted along the frontal plane (to the left and right, changing the roll). Further, the listener can move along three spatial dimensions. This also applies to any sound source with a non-omnidirectional directivity, like musical instruments, talkers, or loudspeakers.

Generally, head pitch changes the orientation of the pinnae relative to the source, causing a change in monaural cues, but not necessarily changing the position of the ears, yielding only a minor change in binaural cues. In contrast, head yaw changes the ears position in a diametrically opposed fashion, affecting all interaural cues. Hence, horizontal rotations provide dynamic acoustic cues to distinguish the position of a source in front from one in back, that is, to acoustically resolve the cone of confusion \citep{perrett_available_1995}.

The reasons for moving the head due to the sound are manifold. For example, the reflexive orienting response, i.e., gaze shift combined with head movements (and in some animals combined with pinna movements), allows the listener to orient to the source for further inspection \citep{sokolov_orienting_2001}. While this reflexive mechanism has been investigated often in the past, not much is known about intentional listeners’ head movements in acoustic environments \citep[e.g.,][]{leung_head_2016}. Nevertheless, head movements help in localizing \citep{mcanally_sound_2014} and externalizing \citep{brimijoin_contribution_2013} sounds as well as tracking auditory targets \citep{leung_head_2016} -- tasks requiring a neural basis for the formation of 3-D space. 

\section{Neurophysiology: coding auditory space}\label{Sec:neuro} 

The formation of 3-D space depends not only on the binaural signals but also on the neural system processing them. Auditory spatial perception is formed via neural activities ascending from the auditory nerve to the cortex. This ascending auditory pathway can be anatomically separated into the primary lemniscal pathway and the non-primary (both lemniscal and non-lemniscal) pathways \citep[e.g.,][]{straka_response_2014}. Neurons within the primary pathway process mainly auditory information and project in a tonotopic organization to the auditory cortex. The non-lemniscal pathways are not fully understood yet \citep[e.g.,][]{lee_exploring_2015} as they link a wide constellation of midbrain, cortical, and limbic-related sites, integrating different types of sensory information and providing information about environmental changes even during sleep. The contribution of both pathways to the formation of the auditory space is described here following the ascending neural organization.

\begin{figure}
	\centerline{\includegraphics[width=1.0\textwidth]{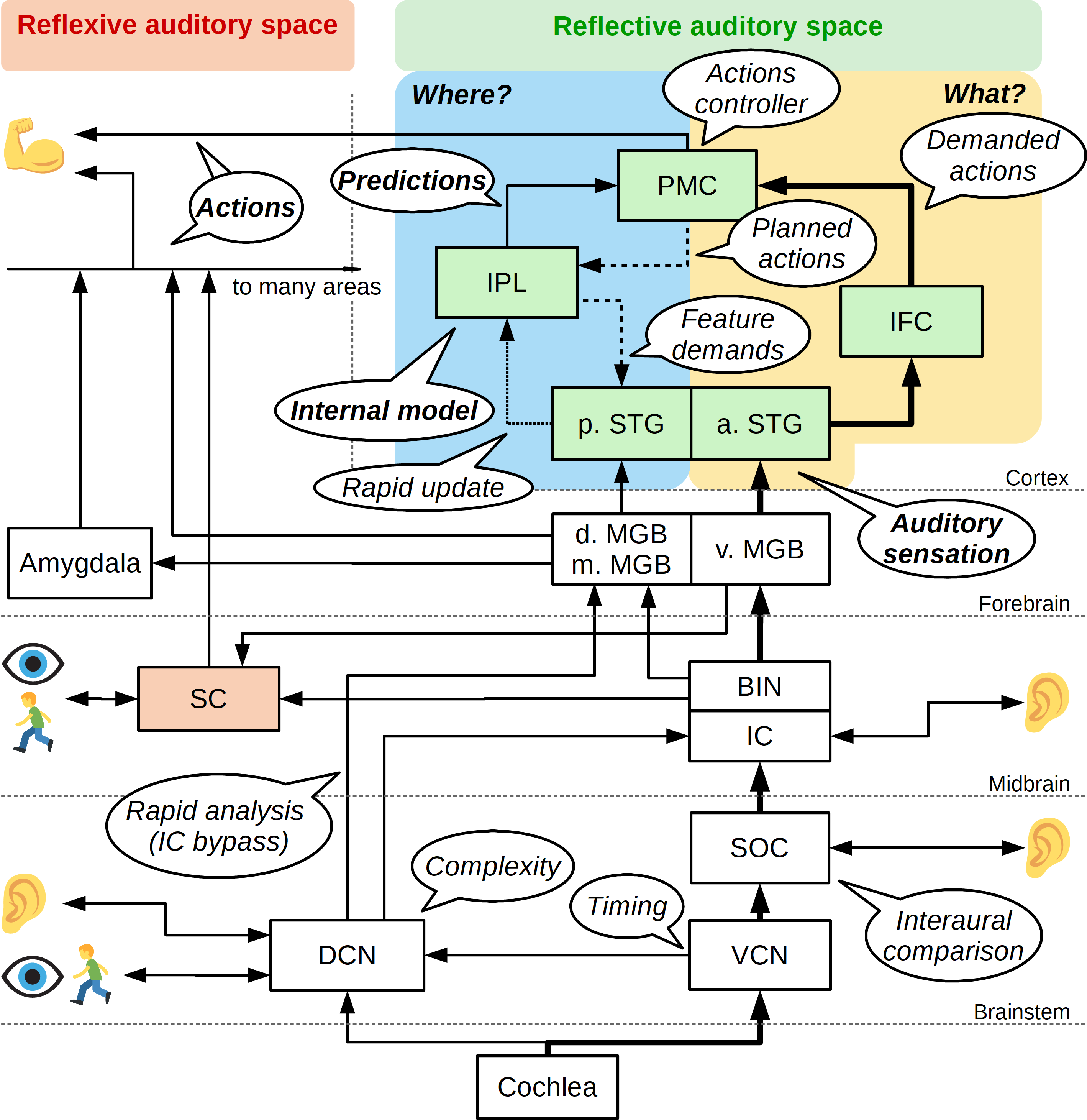}}
	\caption{Ascending neural pathways of human auditory processing relevant in the formation of spatial representation of the environment (see text for more details). \Ear: link to the other contralateral neural side (for more details on the binaural processing, see Pecka et al., this volume.); \Eye: link to the visual system; \Person: link to the somatosensory system; \Arm: link to systems capable of performing actions. Bold lines: the primary auditory pathway. Dashed and doted lines in the "where" pathway: efferent and afferent copies, respectively. Bold text in balloons: correspondence to the cognitive model, see Fig. \ref{Fig:cognition}. }
	\label{Fig:Neuro}
\end{figure}

\subsection{Subcortical pathways: reflexive map of auditory space}
Auditory processing begins in the cochlea where inner hair cells produce neural activity (see Fig.~\ref{Fig:Neuro}). The cochlea is not a simple passive sensor; it is an active sensor whose properties are actively modified by the outer hair cells innervated by efferent connections. The inner hair cells transmit the neural information to spiral ganglion cells whose axons form the auditory nerve (AN, or cochlear nerve). Each of the bilateral ANs projects to the dorsal cochlear nucleus (DCN) and the ventral cochlear nucleus (VCN) in the medulla of the brainstem. Both nuclei are  specialized in decoding certain features of the signal \citep[for an overview, see][Fig. 1]{dehmel_cross-modal_2008}. 

VCN networks receive input from AN fibers tuned to similar (mostly low) frequencies and comprise bushy and octopus cells, which provide the highest temporal precision of any neuron in the brain. The VCN cells project to the contralateral (via the trapezoid body, TB) and ipsilateral superior olivary complexes (SOCs). Given its superior timing properties, the VCN provides a basis for the decoding of timing-critical acoustic features like the ITD.

The DCN is organized differently. Its cells receive inputs not only from the high-frequency AN fibers but also from the VCN and various efferent types of sensory circuits \citep[somatosensory, reticular, vestibular;][]{muniak_tonotopic_2014}. They are similar to those found in the cerebellum, providing evidence for their capability of complex information processing. Indeed, they form tonotopically organized networks that perform nonlinear spectral analysis and separate spectral features according to their expectancy \citep{singla_cerebellum-like_2017}. This analysis helps in forming acoustic cues required for sound-source localization in sagittal planes \citep{may_role_2000}. Further, it may play a role in attenuating body-generated sounds such as vocalization or respiration, and thus may improve the signal-to-noise ratio of the received signal \citep{shore_multisensory_2005}. Interestingly, DCN cells have also (mostly inhibitory) inputs from the contralateral CN, indicating an integration of binaural information already at this very early level of neural processing. The projections of the DCN cells are manifold. Many projections are direct to the inferior colliculus (IC), conveying auditory features including the information required for sound localization based on monaural spectral cues. Other projections are more spread and include a direct path to the thalamus, allowing the cortex to prepare for rapid analysis \citep{anderson_evidence_2006}, and a path to the nucleus reticularis pontis caudalis, critical for triggering the acoustic startle reflex \citep{meloni_dorsal_1998}. 

In the SOC, information is mostly processed in three primary nuclei: the medial superior olive (MSO), the lateral superior olive (LSO), and the medial and lateral nuclei of the TB. The MSO and LSO have a tonotopic organization with bilateral inputs from the VCNs. They form the primary site for the neural computation of ITDs and ILDs (for more details, see Grothe et al. in this volume). Combined with the DCN, at this level of the neural processing, basic auditory spatial features like the ITD, ILD, and monaural spectral features are partially decoded, forming spatial cues available for processing in further neural structures relevant to auditory space. 

In the midbrain, the IC is an obligatory relay for most of the ascending auditory information and combines information from other modalities \citep{gruters_sounds_2012}. Acoustic features are prepared for the formation of auditory objects happening at the next synaptic level. The mammal IC is divided in at least three parts. The central nucleus of the IC is exclusively auditory. It is organized in sheets of isofrequency laminae, each of which receives inputs from multiple different nuclei of the brainstem that permit the decoding of parallel attributes like amplitude and frequency modulation. Binaural interactions appear to be quite complex and nonlinear. Nevertheless, there is strong evidence for the processing of spatial information in all three dimensions. Single neurons have been found showing ITD sensitivity similar to that found in psychophysical experiments. Hence, the lateral angle of a sound can be encoded by means of a single neural path without the need for a population code \citep{skottun_ability_2001}. Single neurons have also been found showing a better temporal coding of reverberant stimuli than that of anechoic stimuli \citep{slama_neural_2015}. This de-reverberation of the signals may not only improve robustness in sound recognition but may also be used to estimate the distance of a sound source by enabling comparisons between direct and reverberant sound energy. Finally, single neurons have been found being sensitive to simple analogs of HRTF-like spectral shapes \citep{davis_auditory_2003}, which also show the presence of cells sensitive to the spectral cues encoding the polar angle of the sound source. The pericentral and external parts of the IC receive ascending inputs from the central IC as well as descending inputs from the somatosensory system, auditory cortex, and other higher brain regions. Many projections are bilateral and thus create numerous feedback loops enabling the integration of the processed auditory information with that arriving from other sensory systems. From the IC, the information is transmitted via the brachium of the IC (BIN) in parallel to the thalamus and superior colliculus (SC). Spatially sensitive neurons in the BIN, mainly projecting to the SC, were found to prefer the natural alignment of spatial cues corresponding to a sound-source position \citep{slee_alignment_2014}.

The SC integrates information from multiple modalities, in particular, auditory spatial with visual and somatosensory information. Here, maps of the visual space, body surface, and auditory space arise from spatially ordered projections from the retina, skin, and acoustic features, respectively. Interestingly, ITDs do not seem to contribute much to that spatial map, instead, spatial tuning seems to mostly rely on spectral ILDs \citep{slee_linear_2013}. The cooperation of neurons combining multimodal afferents and efferents already at this early neural level decreases reaction time, increases stimulus detectability, and enhances perceptual reliability, when information from two modalities is required in a behavioral task \citep{kayser_early_2007}. The SC projects to many motor-related parts of the brain and its organization can be seen as a dynamic map of motor error \citep[pp. 745]{middlebrooks_auditory_2009} with the receptive fields reflecting the deviation between the angle of gaze (including head and eye position) and the target defined by the sensory input (visual, auditory, somatic). As a whole, it plays a critical role in the ability to direct behaviors toward specific objects by helping orient the head and eyes towards something seen and heard \citep{klier_three-dimensional_2003}. Interestingly, on the one hand, these neural and behavioral abilities seem to remain active even in the absence of the cortex \citep{woods_behavior_1964}. On the other hand, dark-reared animals showed a lack of the SC auditory map while still being able to perform auditory-based behavioral tasks \citep{blatt_analysis_1998}. Hence, while the SC clearly demonstrates the formation of a neural topographic map of the auditory space, the SC is most likely not responsible for the creation of the auditory space we are aware of. It is rather a tool created to fast and mostly reflexively react to the environment in form of the orienting response \citep{peck_visual-auditory_1996}. From this, one might conclude that the pathway from SOC via IC to SC forms a reflexive map of the auditory space \citep[see also][]{yao_transformation_2015}, in parallel to the primary ascending path from the IC to the thalamus and cortex.

\subsection{Thalamus and cortex: cognition of auditory space}

The cortex plays an important role in the formation of auditory space because it groups and segregates both spatial and non-spatial perceptual features of an auditory scene into streams that each refer to an auditory object with certain perceptual properties \citep[for review, see][]{micheyl_role_2007}. Before reaching the cortex, the thalamus acts as a hub, relaying information between different subcortical areas and the cerebral cortex. Also, the thalamus has been shown to provide an important integration of other modalities and the preparation of motor responses. Within the thalamus, the auditory information is processed by the medial geniculate body (MGB) of the ventral thalamus. The MGB is further split in dorsal, medial, and ventral sections. 

The dorsal MGB is non-tonotopically organized, its auditory responses can be influenced strongly by non-auditory inputs, and its neurons project mainly to the belt of the auditory cortex \citep{bartlett_organization_2013}. The medial MGB receives various inputs from the BIN and projects broadly across many tonotopic, non-tonotopic, multimodal, and limbic cortical areas, terminating in the cortex and amygdala \citep[for review, see][]{lee_exploring_2015}. There is also a direct pathway from the DCN to the medial MGB, actually bypassing the IC \citep{anderson_evidence_2006}. This connection shows lower latencies than those via the IC, which may be advantageous in the role of the medial MGB in priming the auditory cortex to prepare it for rapid analysis and recruiting the amygdala for rapid emotional responses such as fear. This rapid emotional analysis potentially triggers the startle reflex and auditory looming bias \citep{bach_rising_2008}. It is not clear yet how the dorsal and medial parts of the MGB also contribute to the process of forming auditory objects. 

The ventral MGB has been much better investigated; it processes mostly auditory information, is driven by projections from the central IC, and forms a tonotopically organized relay of binaural (most) and contralateral-only (few) information \citep{lee_topography_2010}. The outputs of the ventral MGB mostly project to the primary auditory cortex and the rostral auditory area where the acoustic features are further processed to build auditory objects.

When finally reaching the cortex, the auditory information is spread among various regions. The most involved regions are the superior temporal gyrus (STG, including the auditory cortex, AC), inferior frontal cortex (IFC), inferior parietal lobe (IPL), and premotor cortex (PMC). There is strong evidence for the existence of two largely segregated processing paths, forming the dual-pathway model of auditory cortical processing, each of them subserving the two main functions of hearing: "what", the identification of auditory objects (e.g., recognition of speech); and "where", the processing of motion and spatial properties of objects \citep[e.g., sound localization;][]{rauschecker_mechanisms_2000}.

The "what" pathway follows the antero-lateral route of the AC, which includes the primary auditory cortex (A1), a rostral area (R), the lateral and medial belt (including the caudomedial area, CM), and the parabelt. Moving along the ascending route within the A1, a transition from the representation of acoustic features (e.g., response to pure tones) via perceptual features (e.g., pitch and timbre) to category representation (e.g., auditory objects) happens. Beginning in the A1, pure-tone sensitive neurons receive inputs from the ventral MBG and are mostly tonotopically organized. A topographic representation of the auditory space, like that found in the SC, is mostly missing in the A1. In contrast, responses of some single neurons vary according to the sound-source direction and result in a 360$^\circ$-like panoramic representation of the space by firing rate \citep{middlebrooks_auditory_2009}, that is, a neural code for spatial information. The “what” pathway ascends further to the IFC via other processing areas within the antero-lateral STG. These and similar connections allow for further processing of non-spatial properties of auditory objects. From the IFC, this information is transformed into articulatory or motor representations in the PMC. 

In contrast, the "where" pathway is highly involved in processing the spatial information retrieved from the auditory stream \citep[for review, see ][]{rauschecker_expanded_2011}. While its exact role is still being debated, its processing involves the separation, location, trajectory, and temporal context of a sound \citep{ahveninen_psychophysics_2014}. The active regions of the "where" pathway are quite diverse; they seem to follow a postero-dorsal non-primary route, with projections from medial and dorsal MGB via the posterior STG (including the planum temporale, PT, that is, the superior surface of the STG) to the IPL and dorsal and ventral areas in the PMC \citep{rauschecker_mechanisms_2000}. The spatial information including ITD and binaural coherence seems to be encoded by a population code found in various areas of the posterior STG \citep{miller_populations_2009}. This information is further projected to the IPL, that has been found to be strongly involved in the processing of the spatial auditory information \citep{arnott_assessing_2004}. From the IPL, the spatial features are integrated with information from other sensory modalities and further projected to the PMC. 

Hence, the PMC is activated by the "what" pathway (via IFC, see bold lines in Fig.~\ref{Fig:Neuro}) and also modulated by the "where" pathway (via IPL). The modulation by the IPL corresponds to a feed-forward system consisting of an internal predictive model of the environment located in the PMC and updated by the multimodal sensory information ascending from the IPL (see Fig.~\ref{Fig:Neuro}). This allows a quick update of the motor system to the new sensory situation.

The PMC reacting to predictions based on sensory information is further underlined by findings showing that the PMC is not only involved during acoustic stimulation but also during musical imagery \citep{leaver_brain_2009} where it is responsible for assembling the motor patterns for the potential production of sound sequences. Efferent feedback from the PMC about planned motor actions (dashed line in Fig.~\ref{Fig:Neuro}) together with the fast and temporally precise afferent projections from the posterior STG (i.e., an efferent copy, dotted line in Fig.~\ref{Fig:Neuro}), allow the IPL to compare the spatial auditory information with the predictive motor states, to decide about the required adjustments of the internal model, and to minimize the surprise (compare Sec.~\ref{SubSec:Ill-posed}). Further efferent projections to the STG (dashed line in Fig.~\ref{Fig:Neuro}), on the other hand, allow for modulating the process of feature extraction in the STG according to the changing feature demands \citep[for more details, see][]{rauschecker_expanded_2011}. 

These considerations show that many cortical regions are involved in processing features of the auditory space and there is no clear evidence for a single region representing the auditory space in the cortex per se. The spatial information is transmitted via the firing rate of the neural population. This process is further modulated by vision (see Mendonca et al., in this volume), proprioception \citep{genzel_dependence_2016}, and in particular attention (e.g., Fels et al., in this volume), indicating its reflective nature in the formation of the auditory space. This is in contrast to the well-localized but reflexive map of the auditory space found at the level of the SC.

The summary presented in this section is just a simplification of all the reflexive and reflective processes involved. The human brain is an analog, high-dimensional, recurrent, nonlinear, stochastic, and dynamic system \citep{dotov_putting_2014}. At the end of the day, these processes form the perception that allows humans to complete various spatial tasks. In the following section, we describe psychophysical spatial tasks, all of them demonstrating the ability to utilize our understanding of the 3-D auditory world.

\section{Psychophysics: listener’s abilities given the perceived auditory space}\label{Sec:Psycho}
Spatial auditory cues facilitate both reflexive and reflective behavior. Human psychoacoustic studies usually imply reflective behavioral tasks, while reflexive behavior, if not targeted explicitly, is more commonly tested in populations with limited cognitive abilities such as infants. In this section, we thus review reflective spatial abilities in humans, organized by increasing cognitive complexity.

\subsection{Sound localization}
Sound localization describes the (reflective) ability to estimate the spatial position of the sound source \citep[for review, see][]{middlebrooks_sound_2015}. Sound localization experiments are often conducted in anechoic environments or free-field simulated virtual space and present target sounds either via loudspeakers or via headphones in virtual auditory spaces simulated by filtering with HRTFs. Considering normal-hearing conditions, the major cues are conceptually different for each dimension of the interaural-polar coordinate system. Alterations of the acoustic environment may change the informative character of a certain cue and consequently affect its perceptual weight \citep{keating_complementary_2015}. Neural plasticity not only enables such context-dependent weighting between cues but also enables adapting and extending the set of spatial cues \citep[for review, see][]{mendonca_review_2014}.

Interaural cues facilitate sound localization within the lateral dimension. The duplex theory describes ITD cues being most effective for low-frequency sounds and ILD cues for high-frequency sounds. Localization of broadband sounds in the lateral direction is dominated by low-frequency ITDs \citep{macpherson_listener_2002}. Besides an intermediate frequency range between 2 and 4\,kHz, where robust localization cues are lacking, human listeners are accurate in lateral-angle localization. Minimum audible angles (MAAs) between two successively presented sounds may be as small as 1$^\circ$ for frontal sources, but degrade with increasing laterality to approximately 10$^\circ$ \citep{perrott_minimum_1990}. These thresholds were derived in free-field experiments providing natural combinations of ITD and ILD cues. Tested in isolation, discrimination thresholds of interaural phase differences (leading to ITDs) also increase by an order of magnitude with increasing reference difference \citep[e.g., $2^\circ$ at $0^\circ$ reference increasing to $20^\circ$ at $90^\circ$ reference for a 900\,Hz tone,][]{yost_discriminations_1974}, which closely resembles the observed MAAs. ITD cues result either from the temporal fine structure at frequencies below approximately 1.4\,kHz or from temporal envelopes at higher frequencies. Consistent with the small perceptual weight in the duplex theory, just-noticeable differences (JNDs) for such high-frequency envelope ITDs are at least twice as large as the JNDs at low frequencies \citep{bernstein_enhancing_2002}. ILD JNDs are in the range of 1 dB for high-frequency sounds and depend on amplitude modulation rates. Effects of temporal modulations on ILD JNDs can be explained based on the interaural difference in neural discharge rates with no need for a particular binaural adaptation mechanisms \citep{laback_temporal_2017}. Computational models of lateral-angle localization have a long history and diversity. Recently, a large-scale attempt to systematically investigate these approaches has been initiated \citep{dietz_framework_2017}. The auditory system is at least partly able to adapt to changes in ITD and ILD cues according to visual and audiomotor feedback \citep[e.g.,][]{trapeau_adaptation_2015}. Adaptation aftereffects were only observed in short-term but not long-term studies \citep{trapeau_adaptation_2015}, indicating that different mechanisms are involved.

While spectral-shape cues may be important for lateral-angle localization in monaural but not binaural hearing \citep{macpherson_listener_2002}, they are crucial for sound localization within the sagittal dimension. Consequently, polar-angle localization requires some bandwidth, especially at high frequencies, in order to achieve high spatial acuity, as indicated by MAAs as small as 4$^\circ$ \citep{perrott_minimum_1990}. Due to redundant spectral information, localization performance is maintained with small uninformative parts in the spectrum. While spectral information limited to frequencies below 16\,kHz seems to be sufficient, limitations down to 8\,kHz cause marked degradations \citep{best_role_2005}. Spectral degradations often result in localization responses biased toward the horizon and led to the concept of “elevation gain” in polar-angle localization \citep{hofman_relearning_1998}. Spectral-shape cues are arguably processed within monaural pathways and thus are often referred to as “monaural spectral cues” although information from both ears is combined following a spatially systematic binaural weighting scheme \citep{macpherson_binaural_2007}: The contralateral ear contributes less with increasing lateral eccentricity where the head shadow will naturally degrade the energy ratio between the target signal and diffuse background noise. Due to the monaural processing, localization performance can be affected by frequency modulations in the stimulus spectrum. Template-based models are able to explain these interactions and show how monaural spectral cues extracted on the basis of tonotopic gradients are rather independent of naturally common low-frequency modulations in the source spectrum \citep{baumgartner_modeling_2014}. Listeners are able to learn new spectral-shape cues \citep{majdak_effect_2013,hofman_relearning_1998} and use them simultaneously with previously acquired cues \citep{trapeau_fast_2016}. Localization performance along sagittal planes is particularly listener-specific, but this variation is hardly explained by only considering the acoustic factor of listener-specifc HRTFs, suggesting large inter-individual differences in how efficient the auditory system is able to  utilize the acoustic information \citep{majdak_acoustic_2014,baumgartner_modeling_2016}.  

To estimate the distance of a source, listeners have access to a broad variety of acoustic cues like sound intensity, reverberation characteristics (often quantified by the direct-to-reverberant energy ratio), near-field ILDs, the shape of the stimulus spectrum, and others \citep[for review, see][]{kolarik_auditory_2016}. The relative perceptual relevance of these cues and their underlying neural codes are the subject of debate and are most probably dependent on the context \citep[for a review, see][]{hladek_temporal_2017}. Recent studies suggest that the amount of temporal ILD fluctuations and amplitude modulation depth likely represent reverberation-related cues \citep{catic_role_2015}. Moreover, spectral-shape cues can affect distance perception \citep{baumgartner_asymmetries_2017} and familiarization to non-individualized spectral-shape cues can improve distance perception \citep{mendonca_learning_2013}.

A special case of distance perception concerns distances closer than physically plausible, namely, inside the listener’s head. Sounds are naturally perceived outside the head (externalized) whereas sounds reproduced with headphones or hearing-assistive devices are often perceived to originate from inside the head \citep[internalized,][]{noble_effects_2006}. Sound externalization is not directly related to the playback device, as free-field signals can be internalized as well \citep{brimijoin_contribution_2013}. Most psychoacoustic studies investigated sound externalization either with discrimination tasks between real and virtual sources and/or distance ratings \citep{hartmann_externalization_1996}. Although perceived distance may be only one of many cues to discriminate between virtual and real sources, the similarity of findings for both paradigms suggests that distance perception is a crucial component. One could think that the only reason for using the term “sound externalization” instead of “distance” is that the percept of sound internalization is an available option. At first glance, however, there seem to be some contradictions between studies focusing on sound externalization and distance perception. For example, decreasing low-frequency ILDs were associated with increasing distance \citep{brungart_auditory_1999} whereas ILDs gradually removed from the low-frequency partials of a harmonic complex \citep{hartmann_externalization_1996} or decreased from broadband speech \citep{brimijoin_contribution_2013} were associated with reduced sound externalization. It seems as if sound internalization is a default state similar to the concept of elevation gain in the case of missing or implausible cues available to the auditory system because no plausible model of the environment can be established to create an external state of affairs.

Head rotation causes dynamic cues particularly effective in resolving front-back confusions \citep{mcanally_sound_2014}. Self-motion affects the actual sound location, but this interaction is successfully compensated by mechanisms responsible for building an allocentric frame of reference \citep{yost_judging_2015}. Listeners are able to create such allocentric spatial representations also without visual information \citep{viaud-delmon_ear_2014}. In contrary to self-motion, listeners are sensitive to source motion and are able to detect movement angles as small as 2$^\circ$ depending on source velocity, stimulus duration, and bandwidth \citep[for review, see][]{carlile_perception_2016}.

\subsection{From spatial impression to presence}
Auditory perception in reverberant spaces like rooms or concert halls is multidimensional \citep{cerda_room_2009}. Listeners are still able to localize the direct sound despite the presence of early reflections, which technically might be interpreted as separate auditory objects. In a room, it is not only that humans are able to integrate the divergent spatial cues into a single auditory object, but also to a certain level, they are even not able to perceive the reflected copies of the direct sound. This ability to suppress early reflections and actually to not perceive them as echoes is referred to as the precedence effect (for review, see Clapp and Seeber, this volume). The delay of the reflections relevant for the precedence effect depends on stimulus; it is around 5~ms for short clicks, and can be up to 30~ms for complex stimuli like speech.

Although we simply perceive a single auditory object even in the presence of its acoustic reflections, the presence of reverberation introduces other spatial effects, which have been summarized as spatial impression or spaciousness \citep{kuhl_spaciousness_1978}. Two main components have been found: apparent source width (ASW) and listener envelopment \citep[LEV,][]{bradley_objective_1995}. 

ASW describes the spatial compactness of a sound event perceived by the listener. In headphone experiments, the main cue for the compactness of the perceived sound is the IACC: If it decreases, the sound is perceived as a wider image \citep{blauert_spatial_1986}. Interestingly, for narrow-band sounds, listeners are extremely sensitive to the deviation of a perfectly coherent signal; they can easily discriminate between signals with an IACC of 1 and 0.99 \citep{gabriel_interaural_1981}. The current explanation for such a high sensitivity is that even in a slightly incoherent signal, large instantaneous ITD and ILD fluctuations occur, which can easily be detected by the auditory system. The perceptual consequence of the fluctuations depends on their duration, bandwidth, center frequency, and stimulus sound level \citep{goupell_interaural_2006}. ASW is, however, not much affected between IACCs of 1 and 0.99 but instead gradually declines with the IACC \citep{whitmer_measuring_2013}. For the extreme case of interaural decorrelation (IACC of zero), the perceived auditory image splits into two objects appearing in the left and the right ears, respectively. 

In reverberant environments where multiple reflections overlap the direct sound, the ASW has been found to be determined by the lateral energy fraction and IACC of the early arriving sound field, that is, within the first 80\,ms of the BRIR \citep{okano_relations_1998}. Deviations of that IACC from one contributes to the perceived quality of concert halls and has been termed the binaural quality index (BQI) of room acoustics. Interestingly, as the BQI increases, the low-frequency ITDs are more likely to be disturbed. Thus, spatial hearing in reverberant situations seems to rely more on high-frequency ITD cues (transmitted in the signal envelope) than on low-frequency ITD cues \citep{ruggles_why_2012}. 

LEV describes how immersed in the sound field the listener feels. It depends on the level, direction of arrival, and temporal properties of later (after 80 ms) arriving reflections \citep{bradley_objective_1995}. Other studies show that late sound arriving from the side, overhead, and behind the listener correlates strongly with LEV, suggesting that the LEV can be distinguished from the late sound having non-lateral components \citep{furuya_arrival_2001}. The late sound arriving from behind and above the listener seem to be important as well \citep{morimoto_role_2001}, showing the relevance of an accurate formation of 3-D auditory space for the perception of rooms. 

The ASW and LEV can be predicted with a model based on the BRIRs and just noticable differences of the IACC \citep{klockgether_model_2014}. The overall perceived quality of concert halls can be quantitatively explained by additional consideration of the reverberation time \citep{cerda_subjective_2015}. 

While both, the ASW and LEV seem to be the major parameters describing the spatial impression of a room, they can both be seen as parts of a larger concept widely used in the context of virtual environments (VE). Here, immersion, as a measure of the psychological sensation of being surrounded \citep{begault_headphone_1998}, integrates the objectively derived LEV and ASW, and further extends to “a psychological state characterized by perceiving oneself to be enveloped by, included in, and interacting with an environment that provides a continuous stream of stimuli and experiences” \citep{witmer_measuring_1998}. In that context, responses to a given level of immersion have been defined as presence, a measure of the psychological sensation of being elsewhere \citep[e.g.,][]{slater_note_2003}. Both immersion and presence are important for the quality of experience in VE systems \citep{moller_quality_2014}. For example, in headphone-based VE systems, immersion can be enhanced with the use of listener-specific HRTFs \citep{hale_virtual_2014}, being in line with studies showing that our auditory system prefers the natural combination of ITDs and ILDs \citep{salminen_integrated_2015}. Here, immersion seems to be more easily conveyed via audio than with vision because audio operates all around the listener even outside the listener's field of view and without exploratory head movements. Immersion and presence seem to be very related attributes and the underlying mechanisms are not fully understood yet \citep[for review, see][]{gaggioli_quality_2003}.

\subsection{Other spatially related tasks}
Spatial hearing also improves tasks not directly related to the formation of the 3-D space. A famous example is the cocktail-party effect which describes the ability to focus on and thus improve the intelligibility of a particular talker in a multi-talker environment \citep{bronkhorst_cocktail-party_2015}. In such a task, a perception of the spatial world is not required per se, however, the benefit of spatial separation of maskers from the target, also called spatial unmasking or release from masking, is clear and has been considered in models predicting speech intelligibility from binaural signals in many situations \citep[e.g.,][]{lavandier_binaural_2012}. Spatial unmasking can further reduce cognitive load in conditions providing similar speech intelligibility \citep[e.g.,][]{andeol_spatial_2017}. Spatial attention, that is, knowing “where” to focus, further modulates the effect of spatial unmasking on a very listener-specific basis \citep{oberfeld_individual_2016}. 

Note that spatial unmasking is not only limited to spatially separated targets and/or noise. Improved speech intelligibility has also been shown in listeners once they have adapted to the acoustics of the listening room \citep{brandewie_prior_2010} indicating that while our auditory system is able to adapt to reverberant spaces, the “noise source” in spatial unmasking can be both additional sound sources and acoustic reflections of the same source. 

Spatial hearing contributes to other, less-known non-spatial tasks. For example, spatial impression can increase the emotional impact of orchestra music by enhancing musical dynamics \citep{patynen_concert_2016}. Looming bias, that is, the phenomenon that approaching sounds are more salient than receding sounds, can be significantly mediated by sound externalization created by the acoustic spatial pinna features alone \citep{baumgartner_asymmetries_2017}. These and similar findings underline the relevance of the formation of 3-D auditory space in our everyday life.
  
\section{Conclusions}
The formation of the auditory space is one of the cognitive processes required to understand and interact with our environment -- by the sensation of sound. In that process, the auditory system has to cope with ephemeral acoustic information about the auditory objects around us. The spatial information, conveyed by the binaural signals, is encoded by interaural and monaural features, along various temporal ranges. Our neural auditory system then creates two representations of the auditory space: a topographically structured neural network in the superior colliculus, capable of triggering quick reflexive reactions; and a reflective cortical representation, encoded by neural populations, capable of modulating other cognitive processes by means of attention. The reflective representation allows us to consciously perceive the auditory space and perform spatial tasks. 

Many concepts have been proposed for cognitive processes involved in the formation of the auditory space. Our interaction with the environment can be seen as a feedforward system with the internal model anticipating the external (or distal) state of affairs. Feedback coming from hearing and other senses allows us to compensate for any deviations to the predictions. Given the ambiguity in the estimation of the external state of affairs from the limited binaural information, the free-energy principle, also known as the active inference, seems to be a promising approach to explain how cognition restricts itself to a limited number of states. Further progress in the development of mathematical methods for solving ill-posed problems and of experimental methods combining psychophysics with neurophysiology will help to improve our understanding of the formation of the auditory space in the future. This will lead to advances in many technical applications like hearing aids driven by spatial attention, listener-specific virtual acoustics, and dynamic sound reproduction systems. 
  
\section*{Acknowledgement}
We would like to thank S. Clapp and B. Seeber for their valuable comments and suggestions. Supported by the Austrian Science Fund (FWF, J 3803-N30).


\begin{thebibliography}{153}
	\def\enquote#1{``#1,''}
	\def\plainquote#1{``#1''}
	\expandafter\ifx\csname natexlab\endcsname\relax\def\natexlab#1{#1}\fi
	\providecommand{\dourl}[1]{\href{http://#1}{\nolinkurl{#1}}}
	\providecommand{\bibinfo}[2]{#2}
	\providecommand{\noopsort}[1]{}
	\providecommand{\switchargs}[2]{#2#1}
	\def\eatspace #1{#1}
	\providecommand{\dodoi}[1]{doi: \href{http://dx.doi.org/#1}{\nolinkurl{#1}}}
	
	\bibitem[{Ahveninen \emph{et~al.}(2014)Ahveninen, Kop\v{c}o, and
		J\"{a}\"{a}skel\"{a}inen}]{ahveninen_psychophysics_2014}
	\bibinfo{author}{Ahveninen, J.}, \bibinfo{author}{Kop\v{c}o, N.},  and
	\bibinfo{author}{J\"{a}\"{a}skel\"{a}inen, I.~P.}
	(\textbf{\bibinfo{year}{2014}}).
	\enquote{\bibinfo{title}{\href{http://dx.doi.org/10.1016/j.heares.2013.07.008}{Psychophysics
				and neuronal bases of sound localization in humans}}}
	\bibinfo{journal}{Hearing Research} \textbf{307}, \bibinfo{pages}{86--97}.
	
	\bibitem[{Aitchison and Lengyel(2017)}]{aitchison_or_2017}
	\bibinfo{author}{Aitchison, L.},  and \bibinfo{author}{Lengyel, M.}
	(\textbf{\bibinfo{year}{2017}}).
	\enquote{\bibinfo{title}{\href{http://dx.doi.org/10.1016/j.conb.2017.08.010}{With
				or without you: predictive coding and {Bayesian} inference in the brain}}}
	\bibinfo{journal}{Curr Opin Neurobiol} \textbf{46},
	\bibinfo{pages}{219--227}.
	
	\bibitem[{Algazi \emph{et~al.}(2001)Algazi, Avendano, and
		Duda}]{algazi_elevation_2001}
	\bibinfo{author}{Algazi, V.~R.}, \bibinfo{author}{Avendano, C.},  and
	\bibinfo{author}{Duda, R.~O.} (\textbf{\bibinfo{year}{2001}}).
	\enquote{\bibinfo{title}{\href{http://view.ncbi.nlm.nih.gov/pubmed/11303925}{Elevation
				localization and head-related transfer function analysis at low
				frequencies}}} \bibinfo{journal}{J Acoust Soc Am} \textbf{109}(3),
	\bibinfo{pages}{1110--22}.
	
	\bibitem[{And\'{e}ol \emph{et~al.}(2017)And\'{e}ol, Suied, Scannella, and
		Dehais}]{andeol_spatial_2017}
	\bibinfo{author}{And\'{e}ol, G.}, \bibinfo{author}{Suied, C.},
	\bibinfo{author}{Scannella, S.},  and \bibinfo{author}{Dehais, F.}
	(\textbf{\bibinfo{year}{2017}}).
	\enquote{\bibinfo{title}{\href{http://dx.doi.org/10.1007/s10162-016-0611-7}{The
				{Spatial} {Release} of {Cognitive} {Load} in {Cocktail} {Party} {Is}
				{Determined} by the {Relative} {Levels} of the {Talkers}}}}
	\bibinfo{journal}{J Assoc Res Otolaryngol} \bibinfo{pages}{1--8}.
	
	\bibitem[{Anderson \emph{et~al.}(2006)Anderson, Malmierca, Wallace, and
		Palmer}]{anderson_evidence_2006}
	\bibinfo{author}{Anderson, L.~A.}, \bibinfo{author}{Malmierca, M.~S.},
	\bibinfo{author}{Wallace, M.~N.},  and \bibinfo{author}{Palmer, A.~R.}
	(\textbf{\bibinfo{year}{2006}}).
	\enquote{\bibinfo{title}{\href{http://dx.doi.org/10.1111/j.1460-9568.2006.04930.x}{Evidence
				for a direct, short latency projection from the dorsal cochlear nucleus to
				the auditory thalamus in the guinea pig}}} \bibinfo{journal}{Eur J Neurosci}
	\textbf{24}(2), \bibinfo{pages}{491--498}.
	
	\bibitem[{Andreopoulou and Katz(2017)}]{andreopoulou_identification_2017}
	\bibinfo{author}{Andreopoulou, A.},  and \bibinfo{author}{Katz, B. F.~G.}
	(\textbf{\bibinfo{year}{2017}}).
	\enquote{\bibinfo{title}{\href{http://dx.doi.org/10.1121/1.4996457}{Identification
				of perceptually relevant methods of inter-aural time difference estimation}}}
	\bibinfo{journal}{J Acoust Soc Am} \textbf{142}(2),
	\bibinfo{pages}{588--598}.
	
	\bibitem[{Arnal \emph{et~al.}(2015)Arnal, Flinker, Kleinschmidt, Giraud, and
		Poeppel}]{arnal_human_2015}
	\bibinfo{author}{Arnal, L.}, \bibinfo{author}{Flinker, A.},
	\bibinfo{author}{Kleinschmidt, A.}, \bibinfo{author}{Giraud, A.-L.},  and
	\bibinfo{author}{Poeppel, D.} (\textbf{\bibinfo{year}{2015}}).
	\enquote{\bibinfo{title}{\href{http://dx.doi.org/10.1016/j.cub.2015.06.043}{Human
				{Screams} {Occupy} a {Privileged} {Niche} in the {Communication}
				{Soundscape}}}} \bibinfo{journal}{Current Biology} \textbf{25}(15),
	\bibinfo{pages}{2051--2056}.
	
	\bibitem[{Arnott \emph{et~al.}(2004)Arnott, Binns, Grady, and
		Alain}]{arnott_assessing_2004}
	\bibinfo{author}{Arnott, S.~R.}, \bibinfo{author}{Binns, M.~A.},
	\bibinfo{author}{Grady, C.~L.},  and \bibinfo{author}{Alain, C.}
	(\textbf{\bibinfo{year}{2004}}).
	\enquote{\bibinfo{title}{\href{http://dx.doi.org/10.1016/j.neuroimage.2004.01.014}{Assessing
				the auditory dual-pathway model in humans}}} \bibinfo{journal}{NeuroImage}
	\textbf{22}(1), \bibinfo{pages}{401--408}.
	
	\bibitem[{Awh \emph{et~al.}(2012)Awh, Belopolsky, and
		Theeuwes}]{awh_top-down_2012}
	\bibinfo{author}{Awh, E.}, \bibinfo{author}{Belopolsky, A.~V.},  and
	\bibinfo{author}{Theeuwes, J.} (\textbf{\bibinfo{year}{2012}}).
	\enquote{\bibinfo{title}{\href{http://dx.doi.org/10.1016/j.tics.2012.06.010}{Top-down
				versus bottom-up attentional control: a failed theoretical dichotomy}}}
	\bibinfo{journal}{Trends Cogn Sci} \textbf{16}(8), \bibinfo{pages}{437--443}.
	
	\bibitem[{Bach \emph{et~al.}(2008)Bach, Sch\"{a}chinger, Neuhoff, Esposito,
		Di~Salle, Lehmann, Herdener, Scheffler, and Seifritz}]{bach_rising_2008}
	\bibinfo{author}{Bach, D.~R.}, \bibinfo{author}{Sch\"{a}chinger, H.},
	\bibinfo{author}{Neuhoff, J.~G.}, \bibinfo{author}{Esposito, F.},
	\bibinfo{author}{Di~Salle, F.}, \bibinfo{author}{Lehmann, C.},
	\bibinfo{author}{Herdener, M.}, \bibinfo{author}{Scheffler, K.},  and
	\bibinfo{author}{Seifritz, E.} (\textbf{\bibinfo{year}{2008}}).
	\enquote{\bibinfo{title}{\href{http://dx.doi.org/10.1093/cercor/bhm040}{Rising
				sound intensity: an intrinsic warning cue activating the amygdala}}}
	\bibinfo{journal}{Cereb Cortex} \textbf{18}(1), \bibinfo{pages}{145--150}.
	
	\bibitem[{Bartlett(2013)}]{bartlett_organization_2013}
	\bibinfo{author}{Bartlett, E.~L.} (\textbf{\bibinfo{year}{2013}}).
	\enquote{\bibinfo{title}{\href{http://dx.doi.org/10.1016/j.bandl.2013.03.003}{The
				organization and physiology of the auditory thalamus and its role in
				processing acoustic features important for speech perception}}}
	\bibinfo{journal}{Brain and language} \textbf{126}(1),
	\bibinfo{pages}{29--48}.
	
	\bibitem[{Baumgartner \emph{et~al.}(2014)Baumgartner, Majdak, and
		Laback}]{baumgartner_modeling_2014}
	\bibinfo{author}{Baumgartner, R.}, \bibinfo{author}{Majdak, P.},  and
	\bibinfo{author}{Laback, B.} (\textbf{\bibinfo{year}{2014}}).
	\enquote{\bibinfo{title}{\href{http://dx.doi.org/10.1121/1.4887447}{Modeling
				sound-source localization in sagittal planes for human listeners}}}
	\bibinfo{journal}{J Acoust Soc Am} \textbf{136}(2),
	\bibinfo{pages}{791--802}.
	
	\bibitem[{Baumgartner \emph{et~al.}(2016)Baumgartner, Majdak, and
		Laback}]{baumgartner_modeling_2016}
	\bibinfo{author}{Baumgartner, R.}, \bibinfo{author}{Majdak, P.},  and
	\bibinfo{author}{Laback, B.} (\textbf{\bibinfo{year}{2016}}).
	\enquote{\bibinfo{title}{\href{http://dx.doi.org/10.1177/2331216516662003}{Modeling
				the {Effects} of {Sensorineural} {Hearing} {Loss} on {Sound} {Localization}
				in the {Median} {Plane}}}} \bibinfo{journal}{Trends Hear} \textbf{20},
	\bibinfo{pages}{2331216516662003}.
	
	\bibitem[{Baumgartner \emph{et~al.}(2017)Baumgartner, Reed, T\'{o}th, Best,
		Majdak, Colburn, and Shinn-Cunningham}]{baumgartner_asymmetries_2017}
	\bibinfo{author}{Baumgartner, R.}, \bibinfo{author}{Reed, D.~K.},
	\bibinfo{author}{T\'{o}th, B.}, \bibinfo{author}{Best, V.},
	\bibinfo{author}{Majdak, P.}, \bibinfo{author}{Colburn, H.~S.},  and
	\bibinfo{author}{Shinn-Cunningham, B.} (\textbf{\bibinfo{year}{2017}}).
	\enquote{\bibinfo{title}{\href{http://dx.doi.org/10.1073/pnas.1703247114}{Asymmetries
				in behavioral and neural responses to spectral cues demonstrate the
				generality of auditory looming bias}}} \bibinfo{journal}{Proc Natl Acad Sci}
	\textbf{114}(36), \bibinfo{pages}{9743--9748}.
	
	\bibitem[{Begault \emph{et~al.}(1998)Begault, Ellis, and
		Wenzel}]{begault_headphone_1998}
	\bibinfo{author}{Begault, D.~R.}, \bibinfo{author}{Ellis, S.~R.},  and
	\bibinfo{author}{Wenzel, E.~M.} (\textbf{\bibinfo{year}{1998}}).
	\enquote{\bibinfo{title}{\href{http://www.aes.org/e-lib/browse.cfm?elib=8089}{Headphone
				and {Head}-{Mounted} {Visual} {Displays} for {Virtual} {Environments}}}}
	\bibinfo{publisher}{J Audio Eng Soc}, Vol. 49, pp. \bibinfo{pages}{904--916}.
	
	\bibitem[{Bernstein and Trahiotis(2002)}]{bernstein_enhancing_2002}
	\bibinfo{author}{Bernstein, L.~R.},  and \bibinfo{author}{Trahiotis, C.}
	(\textbf{\bibinfo{year}{2002}}). \enquote{\bibinfo{title}{Enhancing
			sensitivity to interaural delays at high frequencies by using "transposed
			stimuli"}} \bibinfo{journal}{J Acoust Soc Am} \textbf{112}(3),
	\bibinfo{pages}{1026--1036}.
	
	\bibitem[{Best \emph{et~al.}(2005)Best, Carlile, Jin, and van
		Schaik}]{best_role_2005}
	\bibinfo{author}{Best, V.}, \bibinfo{author}{Carlile, S.},
	\bibinfo{author}{Jin, C.},  and \bibinfo{author}{van Schaik, A.}
	(\textbf{\bibinfo{year}{2005}}).
	\enquote{\bibinfo{title}{\href{http://dx.doi.org/10.1121/1.1926107}{The role
				of high frequencies in speech localization}}} \bibinfo{journal}{J Acoust Soc
		Am} \textbf{118}(1), \bibinfo{pages}{353--63}.
	
	\bibitem[{Best \emph{et~al.}(2007)Best, Gallun, Carlile, and
		Shinn-Cunningham}]{best_binaural_2007}
	\bibinfo{author}{Best, V.}, \bibinfo{author}{Gallun, F.~J.},
	\bibinfo{author}{Carlile, S.},  and \bibinfo{author}{Shinn-Cunningham, B.~G.}
	(\textbf{\bibinfo{year}{2007}}). \enquote{\bibinfo{title}{Binaural
			interference and auditory grouping}} \bibinfo{journal}{J Acoust Soc Am}
	\textbf{121}(2), \bibinfo{pages}{1070--1076}.
	
	\bibitem[{Bhatt and Quinn(2011)}]{bhatt_how_2011}
	\bibinfo{author}{Bhatt, R.~S.},  and \bibinfo{author}{Quinn, P.~C.}
	(\textbf{\bibinfo{year}{2011}}).
	\enquote{\bibinfo{title}{\href{http://dx.doi.org/10.1111/j.1532-7078.2010.00048.x}{How
				does {Learning} {Impact} {Development} in {Infancy}? {The} {Case} of
				{Perceptual} {Organization}}}} \bibinfo{journal}{Infancy} \textbf{16}(1),
	\bibinfo{pages}{2--38}.
	
	\bibitem[{Bizley and Cohen(2013)}]{bizley_what_2013}
	\bibinfo{author}{Bizley, J.~K.},  and \bibinfo{author}{Cohen, Y.~E.}
	(\textbf{\bibinfo{year}{2013}}).
	\enquote{\bibinfo{title}{\href{http://dx.doi.org/10.1038/nrn3565}{The what,
				where and how of auditory-object perception}}} \bibinfo{journal}{Nat Rev
		Neurosci} \textbf{14}(10), \bibinfo{pages}{693--707}.
	
	\bibitem[{Blatt \emph{et~al.}(1998)Blatt, von Linstow~Roloff, Withington,
		Macphail, and Riedel}]{blatt_analysis_1998}
	\bibinfo{author}{Blatt, B.}, \bibinfo{author}{von Linstow~Roloff, E.},
	\bibinfo{author}{Withington, D.~J.}, \bibinfo{author}{Macphail, E.~M.},  and
	\bibinfo{author}{Riedel, G.} (\textbf{\bibinfo{year}{1998}}).
	\enquote{\bibinfo{title}{Analysis of the superior colliculus auditory space
			map function in guinea pig behavior}} \bibinfo{journal}{Neurosci Res Commun}
	\textbf{23}(1), \bibinfo{pages}{23--40}.
	
	\bibitem[{Blauert(1997)}]{blauert_spatial_1997}
	\bibinfo{author}{Blauert, J.} (\textbf{\bibinfo{year}{1997}}).
	\emph{\bibinfo{title}{Spatial hearing. {The} {Psychophysics} of {Human}
			{Sound} {Localization}}}, \bibinfo{edition}{revised} ed.
	(\bibinfo{publisher}{The MIT Press}, \bibinfo{address}{Cambridge, MA}).
	
	\bibitem[{Blauert and Lindemann(1986)}]{blauert_spatial_1986}
	\bibinfo{author}{Blauert, J.},  and \bibinfo{author}{Lindemann, W.}
	(\textbf{\bibinfo{year}{1986}}).
	\enquote{\bibinfo{title}{\href{http://dx.doi.org/10.1121/1.393471}{Spatial
				mapping of intracranial auditory events for various degrees of interaural
				coherence}}} \bibinfo{journal}{J Acoust Soc Am} \textbf{79}(3),
	\bibinfo{pages}{806--813}.
	
	\bibitem[{Bradley and Soulodre(1995)}]{bradley_objective_1995}
	\bibinfo{author}{Bradley, J.~S.},  and \bibinfo{author}{Soulodre, G.~A.}
	(\textbf{\bibinfo{year}{1995}}).
	\enquote{\bibinfo{title}{\href{http://dx.doi.org/10.1121/1.413225}{Objective
				measures of listener envelopment}}} \bibinfo{journal}{J Acoust Soc Am}
	\textbf{98}(5), \bibinfo{pages}{2590--2597}.
	
	\bibitem[{Brandewie and Zahorik(2010)}]{brandewie_prior_2010}
	\bibinfo{author}{Brandewie, E.},  and \bibinfo{author}{Zahorik, P.}
	(\textbf{\bibinfo{year}{2010}}).
	\enquote{\bibinfo{title}{\href{http://dx.doi.org/10.1121/1.3436565}{Prior
				listening in rooms improves speech intelligibility}}} \bibinfo{journal}{J
		Acoust Soc Am} \textbf{128}(1), \bibinfo{pages}{291--299}.
	
	\bibitem[{Bregman(1990)}]{bregman_auditory_1990}
	\bibinfo{author}{Bregman, A.~S.} (\textbf{\bibinfo{year}{1990}}).
	\emph{\bibinfo{title}{Auditory scene analysis}},
	\bibinfo{volume}{\textbf{10}} (\bibinfo{publisher}{MIT Press},
	\bibinfo{address}{Cambridge, MA}).
	
	\bibitem[{Brimijoin \emph{et~al.}(2013)Brimijoin, Boyd, and
		Akeroyd}]{brimijoin_contribution_2013}
	\bibinfo{author}{Brimijoin, W.~O.}, \bibinfo{author}{Boyd, A.~W.},  and
	\bibinfo{author}{Akeroyd, M.~A.} (\textbf{\bibinfo{year}{2013}}).
	\enquote{\bibinfo{title}{\href{http://dx.doi.org/10.1371/journal.pone.0083068}{The
				{Contribution} of {Head} {Movement} to the {Externalization} and
				{Internalization} of {Sounds}}}} \bibinfo{journal}{PloS One} \textbf{8}(12),
	\bibinfo{pages}{e83068}.
	
	\bibitem[{Bronkhorst(2015)}]{bronkhorst_cocktail-party_2015}
	\bibinfo{author}{Bronkhorst, A.~W.} (\textbf{\bibinfo{year}{2015}}).
	\enquote{\bibinfo{title}{\href{http://dx.doi.org/10.3758/s13414-015-0882-9}{The
				cocktail-party problem revisited: early processing and selection of
				multi-talker speech}}} \bibinfo{journal}{Atten Percept Psychophys}
	\textbf{77}(5), \bibinfo{pages}{1465--1487}.
	
	\bibitem[{Brungart \emph{et~al.}(1999)Brungart, Durlach, and
		Rabinowitz}]{brungart_auditory_1999}
	\bibinfo{author}{Brungart, D.~S.}, \bibinfo{author}{Durlach, N.~I.},  and
	\bibinfo{author}{Rabinowitz, W.~M.} (\textbf{\bibinfo{year}{1999}}).
	\enquote{\bibinfo{title}{\href{http://dx.doi.org/10.1121/1.427943}{Auditory
				localization of nearby sources. {II}. {Localization} of a broadband source}}}
	\bibinfo{journal}{J Acoust Soc Am} \textbf{106}(4),
	\bibinfo{pages}{1956--1968}.
	
	\bibitem[{Brungart and Rabinowitz(1999)}]{brungart_auditory_1999-1}
	\bibinfo{author}{Brungart, D.~S.},  and \bibinfo{author}{Rabinowitz, W.~M.}
	(\textbf{\bibinfo{year}{1999}}). \enquote{\bibinfo{title}{Auditory
			localization of nearby sources. {Head}-related transfer functions}}
	\bibinfo{journal}{J Acoust Soc Am} \textbf{106}(3),
	\bibinfo{pages}{1465--1479}.
	
	\bibitem[{Carbon(2014)}]{carbon_understanding_2014}
	\bibinfo{author}{Carbon, C.-C.} (\textbf{\bibinfo{year}{2014}}).
	\enquote{\bibinfo{title}{\href{http://dx.doi.org/10.3389/fnhum.2014.00566}{Understanding
				human perception by human-made illusions}}} \bibinfo{journal}{Front Hum
		Neurosci} \textbf{8}, \bibinfo{pages}{566}.
	
	\bibitem[{Carlile and Corkhill(2015)}]{carlile_selective_2015}
	\bibinfo{author}{Carlile, S.},  and \bibinfo{author}{Corkhill, C.}
	(\textbf{\bibinfo{year}{2015}}).
	\enquote{\bibinfo{title}{\href{http://dx.doi.org/10.1038/srep08662}{Selective
				spatial attention modulates bottom-up informational masking of speech}}}
	\bibinfo{journal}{Sci Rep} \textbf{5}(1), \bibinfo{pages}{8662}.
	
	\bibitem[{Carlile and Leung(2016)}]{carlile_perception_2016}
	\bibinfo{author}{Carlile, S.},  and \bibinfo{author}{Leung, J.}
	(\textbf{\bibinfo{year}{2016}}).
	\enquote{\bibinfo{title}{\href{http://dx.doi.org/10.1177/2331216516644254}{The
				{Perception} of {Auditory} {Motion}}}} \bibinfo{journal}{Trends Hear}
	\textbf{20}, \bibinfo{pages}{1--19}.
	
	\bibitem[{Catic \emph{et~al.}(2015)Catic, Santurette, and
		Dau}]{catic_role_2015}
	\bibinfo{author}{Catic, J.}, \bibinfo{author}{Santurette, S.},  and
	\bibinfo{author}{Dau, T.} (\textbf{\bibinfo{year}{2015}}).
	\enquote{\bibinfo{title}{\href{http://dx.doi.org/10.1121/1.4928132}{The role
				of reverberation-related binaural cues in the externalization of speech}}}
	\bibinfo{journal}{J Acoust Soc Am} \textbf{138}(2),
	\bibinfo{pages}{1154--1167}.
	
	\bibitem[{Cerd\'{a} \emph{et~al.}(2015)Cerd\'{a}, Gim\'{e}nez, Cibri\'{a}n,
		Gir\'{o}n, and Zamarre\~{n}o}]{cerda_subjective_2015}
	\bibinfo{author}{Cerd\'{a}, S.}, \bibinfo{author}{Gim\'{e}nez, A.},
	\bibinfo{author}{Cibri\'{a}n, R.}, \bibinfo{author}{Gir\'{o}n, S.},  and
	\bibinfo{author}{Zamarre\~{n}o, T.} (\textbf{\bibinfo{year}{2015}}).
	\enquote{\bibinfo{title}{\href{http://dx.doi.org/10.1121/1.4906263}{Subjective
				ranking of concert halls substantiated through orthogonal objective
				parameters}}} \bibinfo{journal}{J Acoust Soc Am} \textbf{137}(2),
	\bibinfo{pages}{580--584}.
	
	\bibitem[{Cerd\'{a} \emph{et~al.}(2009)Cerd\'{a}, Gim\'{e}nez, Romero,
		Cibri\'{a}n, and Miralles}]{cerda_room_2009}
	\bibinfo{author}{Cerd\'{a}, S.}, \bibinfo{author}{Gim\'{e}nez, A.},
	\bibinfo{author}{Romero, J.}, \bibinfo{author}{Cibri\'{a}n, R.},  and
	\bibinfo{author}{Miralles, J.} (\textbf{\bibinfo{year}{2009}}).
	\enquote{\bibinfo{title}{\href{http://dx.doi.org/10.1016/j.apacoust.2008.01.001}{Room
				acoustical parameters: {A} factor analysis approach}}} \bibinfo{journal}{Appl
		Acoust} \textbf{70}(1), \bibinfo{pages}{97--109}.
	
	\bibitem[{Chandrasekaran \emph{et~al.}(2014)Chandrasekaran, Koslov, and
		Maddox}]{chandrasekaran_toward_2014}
	\bibinfo{author}{Chandrasekaran, B.}, \bibinfo{author}{Koslov, S.~R.},  and
	\bibinfo{author}{Maddox, W.~T.} (\textbf{\bibinfo{year}{2014}}).
	\enquote{\bibinfo{title}{\href{http://dx.doi.org/10.3389/fpsyg.2014.00825}{Toward
				a dual-learning systems model of speech category learning}}}
	\bibinfo{journal}{Front Psychol} \textbf{5}, \bibinfo{pages}{825}.
	
	\bibitem[{Curtis and D'Esposito(2003)}]{curtis_success_2003}
	\bibinfo{author}{Curtis, C.~E.},  and \bibinfo{author}{D'Esposito, M.}
	(\textbf{\bibinfo{year}{2003}}).
	\enquote{\bibinfo{title}{\href{http://dx.doi.org/10.1162/089892903321593126}{Success
				and {Failure} {Suppressing} {Reflexive} {Behavior}}}} \bibinfo{journal}{J
		Cogn Neurosci} \textbf{15}(3), \bibinfo{pages}{409--418}.
	
	\bibitem[{Davis \emph{et~al.}(2003)Davis, Ramachandran, and
		May}]{davis_auditory_2003}
	\bibinfo{author}{Davis, K.~A.}, \bibinfo{author}{Ramachandran, R.},  and
	\bibinfo{author}{May, B.~J.} (\textbf{\bibinfo{year}{2003}}).
	\enquote{\bibinfo{title}{\href{http://dx.doi.org/10.1007/s10162-002-2002-5}{Auditory
				{Processing} of {Spectral} {Cues} for {Sound} {Localization} in the
				{Inferior} {Colliculus}}}} \bibinfo{journal}{J Assoc Res Otolaryngol}
	\textbf{4}(2), \bibinfo{pages}{148--163}.
	
	\bibitem[{Dehmel \emph{et~al.}(2008)Dehmel, Cui, and
		Shore}]{dehmel_cross-modal_2008}
	\bibinfo{author}{Dehmel, S.}, \bibinfo{author}{Cui, Y.~L.},  and
	\bibinfo{author}{Shore, S.~E.} (\textbf{\bibinfo{year}{2008}}).
	\enquote{\bibinfo{title}{\href{http://dx.doi.org/10.1044/1059-0889(2008/07-0045)}{Cross-{Modal}
				{Interactions} of {Auditory} and {Somatic} {Inputs} in the {Brainstem} and
				{Midbrain} and {Their} {Imbalance} in {Tinnitus} and {Deafness}}}}
	\bibinfo{journal}{Am J Audiol} \textbf{17}(2), \bibinfo{pages}{S193}.
	
	\bibitem[{Dietz \emph{et~al.}(2017)Dietz, Lestang, Majdak, Stern, Marquardt,
		Ewert, Hartmann, and Goodman}]{dietz_framework_2017}
	\bibinfo{author}{Dietz, M.}, \bibinfo{author}{Lestang, J.-H.},
	\bibinfo{author}{Majdak, P.}, \bibinfo{author}{Stern, R.~M.},
	\bibinfo{author}{Marquardt, T.}, \bibinfo{author}{Ewert, S.~D.},
	\bibinfo{author}{Hartmann, W.~M.},  and \bibinfo{author}{Goodman, D. F.~M.}
	(\textbf{\bibinfo{year}{2017}}).
	\enquote{\bibinfo{title}{\href{http://dx.doi.org/10.1016/j.heares.2017.11.010}{A
				framework for testing and comparing binaural models}}} \bibinfo{journal}{Hear
		Res} \textbf{360}, \bibinfo{pages}{92--106}.
	
	\bibitem[{Dokmanic \emph{et~al.}(2013)Dokmanic, Parhizkar, Walther, M~Lu, and
		Vetterli}]{dokmanic_acoustic_2013}
	\bibinfo{author}{Dokmanic, I.}, \bibinfo{author}{Parhizkar, R.},
	\bibinfo{author}{Walther, A.}, \bibinfo{author}{M~Lu, Y.},  and
	\bibinfo{author}{Vetterli, M.} (\textbf{\bibinfo{year}{2013}}).
	\enquote{\bibinfo{title}{\href{http://dx.doi.org/10.1073/pnas.1221464110}{Acoustic
				{Echoes} {Reveal} {Room} {Shape}}}} \bibinfo{journal}{Proc Natl Acad Sci}
	\textbf{110}(30), \bibinfo{pages}{12186--12191}.
	
	\bibitem[{Donoho(2006)}]{donoho_for_2006}
	\bibinfo{author}{Donoho, D.~L.} (\textbf{\bibinfo{year}{2006}}).
	\enquote{\bibinfo{title}{For most large underdetermined systems of linear
			equations the minimal 1-norm solution is also the sparsest solution}}
	\bibinfo{journal}{Commun Pure Appl Math} \textbf{59}(6),
	\bibinfo{pages}{797--829}.
	
	\bibitem[{Dotov(2014)}]{dotov_putting_2014}
	\bibinfo{author}{Dotov, D.~G.} (\textbf{\bibinfo{year}{2014}}).
	\enquote{\bibinfo{title}{\href{http://dx.doi.org/10.3389/fnhum.2014.00795}{Putting
				reins on the brain. {How} the body and environment use it}}}
	\bibinfo{journal}{Front Hum Neurosci} \textbf{8}(795),
	\bibinfo{pages}{1--12}.
	
	\bibitem[{Epstein and Rogers(1995)}]{epstein_perception_1995}
	\bibinfo{editor}{Epstein, W.},  and \bibinfo{editor}{Rogers, S.~J.}, eds.
	(\textbf{\bibinfo{year}{1995}}). Handbook of perception and cognition
	\emph{\bibinfo{title}{Perception of space and motion}}
	(\bibinfo{publisher}{Academic Press}, \bibinfo{address}{San Diego}).
	
	\bibitem[{Francis and Wonham(1976)}]{francis_internal_1976}
	\bibinfo{author}{Francis, B.},  and \bibinfo{author}{Wonham, W.}
	(\textbf{\bibinfo{year}{1976}}).
	\enquote{\bibinfo{title}{\href{http://dx.doi.org/10.1016/0005-1098(76)90006-6}{The
				internal model principle of control theory}}} \bibinfo{journal}{Automatica}
	\textbf{12}(5), \bibinfo{pages}{457--465}.
	
	\bibitem[{Friston(2012)}]{friston_embodied_2012}
	\bibinfo{author}{Friston, K.} (\textbf{\bibinfo{year}{2012}}).
	\enquote{\bibinfo{title}{\href{http://dx.doi.org/10.1007/s10339-012-0519-z}{Embodied
				inference and spatial cognition}}} \bibinfo{journal}{Cognitive Processing}
	\textbf{13}(S1), \bibinfo{pages}{171--177}.
	
	\bibitem[{Friston \emph{et~al.}(2016)Friston, FitzGerald, Rigoli,
		Schwartenbeck, O'Doherty, and Pezzulo}]{friston_active_2016}
	\bibinfo{author}{Friston, K.}, \bibinfo{author}{FitzGerald, T.},
	\bibinfo{author}{Rigoli, F.}, \bibinfo{author}{Schwartenbeck, P.},
	\bibinfo{author}{O'Doherty, J.},  and \bibinfo{author}{Pezzulo, G.}
	(\textbf{\bibinfo{year}{2016}}). \enquote{\bibinfo{title}{Active inference
			and learning}} \bibinfo{journal}{Neurosci Biobehav Rev} \textbf{68},
	\bibinfo{pages}{862--879}.
	
	\bibitem[{Friston \emph{et~al.}(2006)Friston, Kilner, and
		Harrison}]{friston_free_2006}
	\bibinfo{author}{Friston, K.}, \bibinfo{author}{Kilner, J.},  and
	\bibinfo{author}{Harrison, L.} (\textbf{\bibinfo{year}{2006}}).
	\enquote{\bibinfo{title}{\href{http://dx.doi.org/10.1016/j.jphysparis.2006.10.001}{A
				free energy principle for the brain}}} \bibinfo{journal}{J Physiol Paris}
	\textbf{100}(1-3), \bibinfo{pages}{70--87}.
	
	\bibitem[{Furuya \emph{et~al.}(2001)Furuya, Fujimoto, Young~Ji, and
		Higa}]{furuya_arrival_2001}
	\bibinfo{author}{Furuya, H.}, \bibinfo{author}{Fujimoto, K.},
	\bibinfo{author}{Young~Ji, C.},  and \bibinfo{author}{Higa, N.}
	(\textbf{\bibinfo{year}{2001}}).
	\enquote{\bibinfo{title}{\href{http://dx.doi.org/10.1016/S0003-682X(00)00052-9}{Arrival
				direction of late sound and listener envelopment}}} \bibinfo{journal}{Appl
		Acoust} \textbf{62}(2), \bibinfo{pages}{125--136}.
	
	\bibitem[{Gabriel and Colburn(1981)}]{gabriel_interaural_1981}
	\bibinfo{author}{Gabriel, K.~J.},  and \bibinfo{author}{Colburn, H.~S.}
	(\textbf{\bibinfo{year}{1981}}). \enquote{\bibinfo{title}{Interaural
			correlation discrimination: i. bandwidth and level dependence}}
	\bibinfo{journal}{J Acoust Soc Am} \textbf{69}(5),
	\bibinfo{pages}{1394--1401}.
	
	\bibitem[{Gaggioli \emph{et~al.}(2003)Gaggioli, Bassi, and
		Delle~Fave}]{gaggioli_quality_2003}
	\bibinfo{author}{Gaggioli, A.}, \bibinfo{author}{Bassi, M.},  and
	\bibinfo{author}{Delle~Fave, A.} (\textbf{\bibinfo{year}{2003}}).
	\enquote{\bibinfo{title}{Quality of {Experience} in {Virtual}
			{Environments}}} in \emph{\bibinfo{booktitle}{Being there: {Concepts},
			effects and measuerement of user presence in syntetic environments}}, pp.
	\bibinfo{pages}{122--132}.
	
	\bibitem[{Genzel \emph{et~al.}(2016)Genzel, Firzlaff, Wiegrebe, and
		MacNeilage}]{genzel_dependence_2016}
	\bibinfo{author}{Genzel, D.}, \bibinfo{author}{Firzlaff, U.},
	\bibinfo{author}{Wiegrebe, L.},  and \bibinfo{author}{MacNeilage, P.~R.}
	(\textbf{\bibinfo{year}{2016}}).
	\enquote{\bibinfo{title}{\href{http://dx.doi.org/10.1152/jn.00052.2016}{Dependence
				of auditory spatial updating on vestibular, proprioceptive, and efference
				copy signals}}} \bibinfo{journal}{J Neurophysiol} \textbf{116}(2),
	\bibinfo{pages}{765--775}.
	
	\bibitem[{Gordon \emph{et~al.}(1992)Gordon, Webb, and
		Wolpert}]{gordon_one_1992}
	\bibinfo{author}{Gordon, C.}, \bibinfo{author}{Webb, D.~L.},  and
	\bibinfo{author}{Wolpert, S.} (\textbf{\bibinfo{year}{1992}}).
	\enquote{\bibinfo{title}{\href{http://dx.doi.org/10.1090/S0273-0979-1992-00289-6}{One
				cannot hear the shape of a drum}}} \bibinfo{journal}{Bull Am Math Soc}
	\textbf{27}(1), \bibinfo{pages}{134--138}.
	
	\bibitem[{Goupell and Hartmann(2006)}]{goupell_interaural_2006}
	\bibinfo{author}{Goupell, M.~J.},  and \bibinfo{author}{Hartmann, W.~M.}
	(\textbf{\bibinfo{year}{2006}}). \enquote{\bibinfo{title}{Interaural
			fluctuations and the detection of interaural incoherence: {Bandwidth}
			effects}} \bibinfo{journal}{J Acoust Soc Am} \textbf{119}(6),
	\bibinfo{pages}{3971--3986}.
	
	\bibitem[{Gregory(1980)}]{gregory_perceptions_1980}
	\bibinfo{author}{Gregory, R.~L.} (\textbf{\bibinfo{year}{1980}}).
	\enquote{\bibinfo{title}{Perceptions as hypotheses}} \bibinfo{journal}{Philos
		Trans R Soc Lond B Biol Sci} \textbf{290}(1038), \bibinfo{pages}{181--197}.
	
	\bibitem[{Griffiths and Warren(2004)}]{griffiths_what_2004}
	\bibinfo{author}{Griffiths, T.~D.},  and \bibinfo{author}{Warren, J.~D.}
	(\textbf{\bibinfo{year}{2004}}).
	\enquote{\bibinfo{title}{\href{http://dx.doi.org/10.1038/nrn1538}{What is an
				auditory object?}}} \bibinfo{journal}{Nat Rev Neurosci} \textbf{5}(11),
	\bibinfo{pages}{887--892}.
	
	\bibitem[{Grush(2004)}]{grush_emulation_2004}
	\bibinfo{author}{Grush, R.} (\textbf{\bibinfo{year}{2004}}).
	\enquote{\bibinfo{title}{The emulation theory of representation: motor
			control, imagery, and perception}} \bibinfo{journal}{Behav Brain Sci}
	\textbf{27}(3), \bibinfo{pages}{377--396}.
	
	\bibitem[{Gruters and Groh(2012)}]{gruters_sounds_2012}
	\bibinfo{author}{Gruters, K.~G.},  and \bibinfo{author}{Groh, J.~M.}
	(\textbf{\bibinfo{year}{2012}}).
	\enquote{\bibinfo{title}{\href{http://dx.doi.org/10.3389/fncir.2012.00096}{Sounds
				and beyond: multisensory and other non-auditory signals in the inferior
				colliculus}}} \bibinfo{journal}{Front Neural Circuits} \textbf{6}.
	
	\bibitem[{Hartmann \emph{et~al.}(2005)Hartmann, Rakerd, and
		Koller}]{hartmann_binaural_2005}
	\bibinfo{author}{Hartmann, W.~M.}, \bibinfo{author}{Rakerd, B.},  and
	\bibinfo{author}{Koller, A.} (\textbf{\bibinfo{year}{2005}}).
	\enquote{\bibinfo{title}{Binaural {Coherence} in {Rooms}}}
	\bibinfo{journal}{Acta Acust united Ac} \textbf{91}(3),
	\bibinfo{pages}{451--462}.
	
	\bibitem[{Hartmann and Wittenberg(1996)}]{hartmann_externalization_1996}
	\bibinfo{author}{Hartmann, W.~M.},  and \bibinfo{author}{Wittenberg, A.}
	(\textbf{\bibinfo{year}{1996}}). \enquote{\bibinfo{title}{On the
			externalization of sound images}} \bibinfo{journal}{J Acoust Soc Am}
	\textbf{99}(6), \bibinfo{pages}{3678--3688}.
	
	\bibitem[{Helmholtz(1954)}]{helmholtz_sensations_1954}
	\bibinfo{author}{Helmholtz, H.} (\textbf{\bibinfo{year}{1954}}).
	\emph{\bibinfo{title}{On the {Sensations} of {Tone}}}
	(\bibinfo{publisher}{Dover Books on Music}, \bibinfo{address}{NY}).
	
	\bibitem[{Henning(1974)}]{henning_detectability_1974}
	\bibinfo{author}{Henning, G.~B.} (\textbf{\bibinfo{year}{1974}}).
	\enquote{\bibinfo{title}{Detectability of interaural delay in high-frequency
			complex waveforms}} \bibinfo{journal}{J Acoust Soc Am} \textbf{55}(1),
	\bibinfo{pages}{84--90}.
	
	\bibitem[{Hill and Miller(2010)}]{hill_auditory_2010}
	\bibinfo{author}{Hill, K.~T.},  and \bibinfo{author}{Miller, L.~M.}
	(\textbf{\bibinfo{year}{2010}}).
	\enquote{\bibinfo{title}{\href{http://dx.doi.org/10.1093/cercor/bhp124}{Auditory
				{Attentional} {Control} and {Selection} during {Cocktail} {Party}
				{Listening}}}} \bibinfo{journal}{Cereb Cortex} \textbf{20}(3),
	\bibinfo{pages}{583--590}.
	
	\bibitem[{Hinton and Ghahramani(1997)}]{hinton_generative_1997}
	\bibinfo{author}{Hinton, G.~E.},  and \bibinfo{author}{Ghahramani, Z.}
	(\textbf{\bibinfo{year}{1997}}).
	\enquote{\bibinfo{title}{\href{http://dx.doi.org/10.1098/rstb.1997.0101}{Generative
				models for discovering sparse distributed representations}}}
	\bibinfo{journal}{Philos Trans R Soc Lond B Biol Sci} \textbf{352}(1358),
	\bibinfo{pages}{1177--1190}.
	
	\bibitem[{Hl\'{a}dek \emph{et~al.}(2017)Hl\'{a}dek, Tomoriov\'{a}, and
		Kop\v{c}o}]{hladek_temporal_2017}
	\bibinfo{author}{Hl\'{a}dek, ƒ.}, \bibinfo{author}{Tomoriov\'{a}, B.},  and
	\bibinfo{author}{Kop\v{c}o, N.} (\textbf{\bibinfo{year}{2017}}).
	\enquote{\bibinfo{title}{\href{http://dx.doi.org/10.1121/1.5012746}{Temporal
				characteristics of contextual effects in sound localization}}}
	\bibinfo{journal}{J Acoust Soc Am} \textbf{142}(5),
	\bibinfo{pages}{3288--3296}.
	
	\bibitem[{Hofman \emph{et~al.}(1998)Hofman, van Riswick, and van
		Opstal}]{hofman_relearning_1998}
	\bibinfo{author}{Hofman, P.~M.}, \bibinfo{author}{van Riswick, J. G.~A.},  and
	\bibinfo{author}{van Opstal, A.~J.} (\textbf{\bibinfo{year}{1998}}).
	\enquote{\bibinfo{title}{\href{http://dx.doi.org/http://dx.doi.org/10.1038/1633}{Relearning
				sound localization with new ears}}} \bibinfo{journal}{Nature Neurosci}
	\textbf{1}(5), \bibinfo{pages}{417--421}.
	
	\bibitem[{Kayser and Logothetis(2007)}]{kayser_early_2007}
	\bibinfo{author}{Kayser, C.},  and \bibinfo{author}{Logothetis, N.~K.}
	(\textbf{\bibinfo{year}{2007}}).
	\enquote{\bibinfo{title}{\href{http://dx.doi.org/10.1007/s00429-007-0154-0}{Do
				early sensory cortices integrate cross-modal information?}}}
	\bibinfo{journal}{Brain Struct Funct} \textbf{212}(2),
	\bibinfo{pages}{121--132}.
	
	\bibitem[{Keating \emph{et~al.}(2015)Keating, Dahmen, and
		King}]{keating_complementary_2015}
	\bibinfo{author}{Keating, P.}, \bibinfo{author}{Dahmen, J.~C.},  and
	\bibinfo{author}{King, A.~J.} (\textbf{\bibinfo{year}{2015}}).
	\enquote{\bibinfo{title}{\href{http://dx.doi.org/10.1038/nn.3914}{Complementary
				adaptive processes contribute to the developmental plasticity of spatial
				hearing}}} \bibinfo{journal}{Nat Neurosci} \textbf{18}(2),
	\bibinfo{pages}{185--187}.
	
	\bibitem[{Klier(2003)}]{klier_three-dimensional_2003}
	\bibinfo{author}{Klier, E.~M.} (\textbf{\bibinfo{year}{2003}}).
	\enquote{\bibinfo{title}{\href{http://dx.doi.org/10.1152/jn.00763.2002}{Three-{Dimensional}
				{Eye}-{Head} {Coordination} {Is} {Implemented} {Downstream} {From} the
				{Superior} {Colliculus}}}} \bibinfo{journal}{J Neurophysiol} \textbf{89}(5),
	\bibinfo{pages}{2839--2853}.
	
	\bibitem[{Klockgether and van~de Par(2014)}]{klockgether_model_2014}
	\bibinfo{author}{Klockgether, S.},  and \bibinfo{author}{van~de Par, S.}
	(\textbf{\bibinfo{year}{2014}}).
	\enquote{\bibinfo{title}{\href{http://dx.doi.org/10.3813/AAA.918776}{A
				{Model} for the {Prediction} of {Room} {Acoustical} {Perception} {Based} on
				the {Just} {Noticeable} {Differences} of {Spatial} {Perception}}}}
	\bibinfo{journal}{Acta Acust united Ac} \textbf{100},
	\bibinfo{pages}{964--971}.
	
	\bibitem[{Knudsen(2007)}]{knudsen_fundamental_2007}
	\bibinfo{author}{Knudsen, E.~I.} (\textbf{\bibinfo{year}{2007}}).
	\enquote{\bibinfo{title}{\href{http://dx.doi.org/10.1146/annurev.neuro.30.051606.094256}{Fundamental
				{Components} of {Attention}}}} \bibinfo{journal}{Annu Rev Neurosci}
	\textbf{30}(1), \bibinfo{pages}{57--78}.
	
	\bibitem[{Koffka(1935)}]{koffka_principles_1935}
	\bibinfo{author}{Koffka, K.} (\textbf{\bibinfo{year}{1935}}).
	\emph{\bibinfo{title}{Principles of {Gestalt} psychology}}
	(\bibinfo{publisher}{Mimesis Int.}, \bibinfo{address}{London}).
	
	\bibitem[{Kolarik \emph{et~al.}(2016)Kolarik, Moore, Zahorik, Cirstea, and
		Pardhan}]{kolarik_auditory_2016}
	\bibinfo{author}{Kolarik, A.~J.}, \bibinfo{author}{Moore, B. C.~J.},
	\bibinfo{author}{Zahorik, P.}, \bibinfo{author}{Cirstea, S.},  and
	\bibinfo{author}{Pardhan, S.} (\textbf{\bibinfo{year}{2016}}).
	\enquote{\bibinfo{title}{\href{http://dx.doi.org/10.3758/s13414-015-1015-1}{Auditory
				distance perception in humans: a review of cues, development, neuronal bases,
				and effects of sensory loss}}} \bibinfo{journal}{Atten Percept Psychophys}
	\textbf{78}(2), \bibinfo{pages}{373--395}.
	
	\bibitem[{Kondo \emph{et~al.}(2012)Kondo, Pressnitzer, Toshima, and
		Kashino}]{kondo_effects_2012}
	\bibinfo{author}{Kondo, H.~M.}, \bibinfo{author}{Pressnitzer, D.},
	\bibinfo{author}{Toshima, I.},  and \bibinfo{author}{Kashino, M.}
	(\textbf{\bibinfo{year}{2012}}).
	\enquote{\bibinfo{title}{\href{http://dx.doi.org/10.1073/pnas.1112852109}{Effects
				of self-motion on auditory scene analysis}}} \bibinfo{journal}{Proc Natl Acad
		Sci} \textbf{109}(17), \bibinfo{pages}{6775--6780}.
	
	\bibitem[{Kreuzer \emph{et~al.}(2009)Kreuzer, Majdak, and
		Chen}]{kreuzer_fast_2009}
	\bibinfo{author}{Kreuzer, W.}, \bibinfo{author}{Majdak, P.},  and
	\bibinfo{author}{Chen, Z.} (\textbf{\bibinfo{year}{2009}}).
	\enquote{\bibinfo{title}{\href{http://dx.doi.org/doi: 10.1121/1.3177264}{Fast
				multipole boundary element method to calculate head-related transfer
				functions for a wide frequency range}}} \bibinfo{journal}{J Acoust Soc Am}
	\textbf{126}(3), \bibinfo{pages}{1280--1290}.
	
	\bibitem[{Kuhl(1978)}]{kuhl_spaciousness_1978}
	\bibinfo{author}{Kuhl, W.} (\textbf{\bibinfo{year}{1978}}).
	\enquote{\bibinfo{title}{Spaciousness (spatial impression) as a component of
			total room impression}} \bibinfo{journal}{Acustica} \textbf{40}(3),
	\bibinfo{pages}{167--181}.
	
	\bibitem[{Kuhn(1977)}]{kuhn_model_1977}
	\bibinfo{author}{Kuhn, G.~F.} (\textbf{\bibinfo{year}{1977}}).
	\enquote{\bibinfo{title}{\href{http://dx.doi.org/10.1121/1.381498}{Model for
				the interaural time differences in the azimuthal plane}}} \bibinfo{journal}{J
		Acoust Soc Am} \textbf{62}(1), \bibinfo{pages}{157--167}.
	
	\bibitem[{Kulkarni \emph{et~al.}(1999)Kulkarni, Isabelle, and
		Colburn}]{kulkarni_sensitvity_1999}
	\bibinfo{author}{Kulkarni, A.}, \bibinfo{author}{Isabelle, S.~K.},  and
	\bibinfo{author}{Colburn, H.~S.} (\textbf{\bibinfo{year}{1999}}).
	\enquote{\bibinfo{title}{Sensitvity of human subjects to head-related
			transfer-function phase spectra}} \bibinfo{journal}{J Acoust Soc Am}
	\textbf{105 (5)}, \bibinfo{pages}{2821--2840}.
	
	\bibitem[{Laback \emph{et~al.}(2017)Laback, Dietz, and
		Joris}]{laback_temporal_2017}
	\bibinfo{author}{Laback, B.}, \bibinfo{author}{Dietz, M.},  and
	\bibinfo{author}{Joris, P.} (\textbf{\bibinfo{year}{2017}}).
	\enquote{\bibinfo{title}{\href{http://dx.doi.org/10.1121/1.5009563}{Temporal
				effects in interaural and sequential level difference perception}}}
	\bibinfo{journal}{J Acoust Soc Am} \textbf{142}(5),
	\bibinfo{pages}{3267--3283}.
	
	\bibitem[{Lavandier \emph{et~al.}(2012)Lavandier, Jelfs, Culling, Watkins,
		Raimond, and Makin}]{lavandier_binaural_2012}
	\bibinfo{author}{Lavandier, M.}, \bibinfo{author}{Jelfs, S.},
	\bibinfo{author}{Culling, J.~F.}, \bibinfo{author}{Watkins, A.~J.},
	\bibinfo{author}{Raimond, A.~P.},  and \bibinfo{author}{Makin, S.~J.}
	(\textbf{\bibinfo{year}{2012}}).
	\enquote{\bibinfo{title}{\href{http://dx.doi.org/10.1121/1.3662075}{Binaural
				prediction of speech intelligibility in reverberant rooms with multiple noise
				sources}}} \bibinfo{journal}{J Acoust Soc Am} \textbf{131}(1),
	\bibinfo{pages}{218--231}.
	
	\bibitem[{Leaver \emph{et~al.}(2009)Leaver, Van~Lare, Zielinski, Halpern, and
		Rauschecker}]{leaver_brain_2009}
	\bibinfo{author}{Leaver, A.~M.}, \bibinfo{author}{Van~Lare, J.},
	\bibinfo{author}{Zielinski, B.}, \bibinfo{author}{Halpern, A.~R.},  and
	\bibinfo{author}{Rauschecker, J.~P.} (\textbf{\bibinfo{year}{2009}}).
	\enquote{\bibinfo{title}{\href{http://dx.doi.org/10.1523/JNEUROSCI.4921-08.2009}{Brain
				{Activation} during {Anticipation} of {Sound} {Sequences}}}}
	\bibinfo{journal}{J Neurosci} \textbf{29}(8), \bibinfo{pages}{2477--2485}.
	
	\bibitem[{Lee(2015)}]{lee_exploring_2015}
	\bibinfo{author}{Lee, C.~C.} (\textbf{\bibinfo{year}{2015}}).
	\enquote{\bibinfo{title}{\href{http://dx.doi.org/10.3389/fncir.2015.00069}{Exploring
				functions for the non-lemniscal auditory thalamus}}} \bibinfo{journal}{Front
		Neural Circuits} \textbf{9}.
	
	\bibitem[{Lee and Sherman(2010)}]{lee_topography_2010}
	\bibinfo{author}{Lee, C.~C.},  and \bibinfo{author}{Sherman, S.~M.}
	(\textbf{\bibinfo{year}{2010}}).
	\enquote{\bibinfo{title}{\href{http://dx.doi.org/10.1073/pnas.0907873107}{Topography
				and physiology of ascending streams in the auditory tectothalamic pathway}}}
	\bibinfo{journal}{Proc Natl Acad Sci} \textbf{107}(1),
	\bibinfo{pages}{372--377}.
	
	\bibitem[{Leung \emph{et~al.}(2016)Leung, Wei, Burgess, and
		Carlile}]{leung_head_2016}
	\bibinfo{author}{Leung, J.}, \bibinfo{author}{Wei, V.},
	\bibinfo{author}{Burgess, M.},  and \bibinfo{author}{Carlile, S.}
	(\textbf{\bibinfo{year}{2016}}).
	\enquote{\bibinfo{title}{\href{http://dx.doi.org/10.3389/fnins.2015.00493}{Head
				{Tracking} of {Auditory}, {Visual}, and {Audio}-{Visual} {Targets}}}}
	\bibinfo{journal}{Front Neurosci} \textbf{9}.
	
	\bibitem[{Lord Rayleigh~or Strutt(1876)}]{lord_rayleigh_or_strutt_our_1876}
	\bibinfo{author}{Lord Rayleigh~or Strutt, F.} (\textbf{\bibinfo{year}{1876}}).
	\enquote{\bibinfo{title}{Our perception of the direction of a source of
			sound}} \bibinfo{journal}{Proc Musical Association 2}
	\bibinfo{pages}{75--84}.
	
	\bibitem[{Macpherson and Middlebrooks(2002)}]{macpherson_listener_2002}
	\bibinfo{author}{Macpherson, E.~A.},  and \bibinfo{author}{Middlebrooks, J.~C.}
	(\textbf{\bibinfo{year}{2002}}).
	\enquote{\bibinfo{title}{\href{http://dx.doi.org/10.1121/1.1471898}{Listener
				weighting of cues for lateral angle: {The} duplex theory of sound
				localization revisited}}} \bibinfo{journal}{J Acoust Soc Ama}
	\textbf{111}(5), \bibinfo{pages}{2219--2236}.
	
	\bibitem[{Macpherson and Middlebrooks(2003)}]{macpherson_vertical-plane_2003}
	\bibinfo{author}{Macpherson, E.~A.},  and \bibinfo{author}{Middlebrooks, J.~C.}
	(\textbf{\bibinfo{year}{2003}}).
	\enquote{\bibinfo{title}{\href{http://dx.doi.org/10.1121/1.1582174}{Vertical-plane
				sound localization probed with ripple-spectrum noise}}} \bibinfo{journal}{J
		Acoust Soc Am} \textbf{114}(1), \bibinfo{pages}{430--445}.
	
	\bibitem[{Macpherson and Sabin(2007)}]{macpherson_binaural_2007}
	\bibinfo{author}{Macpherson, E.~A.},  and \bibinfo{author}{Sabin, A.~T.}
	(\textbf{\bibinfo{year}{2007}}).
	\enquote{\bibinfo{title}{\href{http://dx.doi.org/10.1121/1.2722048}{Binaural
				weighting of monaural spectral cues for sound localization}}}
	\bibinfo{journal}{J Acoust Soc Am} \textbf{121}(6),
	\bibinfo{pages}{3677--3688}.
	
	\bibitem[{Majdak \emph{et~al.}(2007)Majdak, Balazs, and
		Laback}]{majdak_multiple_2007}
	\bibinfo{author}{Majdak, P.}, \bibinfo{author}{Balazs, P.},  and
	\bibinfo{author}{Laback, B.} (\textbf{\bibinfo{year}{2007}}).
	\enquote{\bibinfo{title}{Multiple exponential sweep method for fast
			measurement of head-related transfer functions}} \bibinfo{journal}{J Audio
		Eng Soc} \textbf{55}, \bibinfo{pages}{623--637}.
	
	\bibitem[{Majdak \emph{et~al.}(2014)Majdak, Baumgartner, and
		Laback}]{majdak_acoustic_2014}
	\bibinfo{author}{Majdak, P.}, \bibinfo{author}{Baumgartner, R.},  and
	\bibinfo{author}{Laback, B.} (\textbf{\bibinfo{year}{2014}}).
	\enquote{\bibinfo{title}{\href{http://dx.doi.org/10.3389/fpsyg.2014.00319}{Acoustic
				and non-acoustic factors in modeling listener-specific performance of
				sagittal-plane sound localization}}} \bibinfo{journal}{Front Psychol}
	\textbf{5}(319), \bibinfo{pages}{1--10}.
	
	\bibitem[{Majdak \emph{et~al.}(2013{\natexlab{a}})Majdak, Carpentier, Nicol,
		Roginska, Suzuki, Watanabe, Wierstorf, Ziegelwanger, and
		Noisternig}]{majdak_spatially_2013}
	\bibinfo{author}{Majdak, P.}, \bibinfo{author}{Carpentier, T.},
	\bibinfo{author}{Nicol, R.}, \bibinfo{author}{Roginska, A.},
	\bibinfo{author}{Suzuki, Y.}, \bibinfo{author}{Watanabe, K.},
	\bibinfo{author}{Wierstorf, H.}, \bibinfo{author}{Ziegelwanger, H.},  and
	\bibinfo{author}{Noisternig, M.}
	(\textbf{\bibinfo{year}{2013}}{\natexlab{a}}).
	\enquote{\bibinfo{title}{Spatially {Oriented} {Format} for {Acoustics}: {A}
			{Data} {Exchange} {Format} {Representing} {Head}-{Related} {Transfer}
			{Functions}}} in \emph{\bibinfo{booktitle}{Proc 134th Conv Audio Eng Soc}},
	\bibinfo{address}{Roma, Italy}, p. \bibinfo{pages}{8880}.
	
	\bibitem[{Majdak \emph{et~al.}(2010)Majdak, Goupell, and
		Laback}]{majdak_3-d_2010}
	\bibinfo{author}{Majdak, P.}, \bibinfo{author}{Goupell, M.~J.},  and
	\bibinfo{author}{Laback, B.} (\textbf{\bibinfo{year}{2010}}).
	\enquote{\bibinfo{title}{\href{http://dx.doi.org/DOI:
				10.3758/APP.72.2.454}{3-{D} localization of virtual sound sources: effects of
				visual environment, pointing method, and training}}} \bibinfo{journal}{Atten
		Percept Psychophys} \textbf{72}(2), \bibinfo{pages}{454--69}.
	
	\bibitem[{Majdak \emph{et~al.}(2013{\natexlab{b}})Majdak, Walder, and
		Laback}]{majdak_effect_2013}
	\bibinfo{author}{Majdak, P.}, \bibinfo{author}{Walder, T.},  and
	\bibinfo{author}{Laback, B.} (\textbf{\bibinfo{year}{2013}}{\natexlab{b}}).
	\enquote{\bibinfo{title}{\href{http://dx.doi.org/10.1121/1.4816543}{Effect of
				long-term training on sound localization performance with spectrally warped
				and band-limited head-related transfer functions}}} \bibinfo{journal}{J
		Acoust Soc Am} \textbf{134}(3), \bibinfo{pages}{2148--2159}.
	
	\bibitem[{Mason \emph{et~al.}(2005)Mason, Brookes, and
		Rumsey}]{mason_frequency_2005}
	\bibinfo{author}{Mason, R.}, \bibinfo{author}{Brookes, T.},  and
	\bibinfo{author}{Rumsey, F.} (\textbf{\bibinfo{year}{2005}}).
	\enquote{\bibinfo{title}{\href{http://dx.doi.org/10.1121/1.1853113}{Frequency
				dependency of the relationship between perceived auditory source width and
				the interaural cross-correlation coefficient for time-invariant stimuli}}}
	\bibinfo{journal}{J Acoust Soc Am} \textbf{117}(3),
	\bibinfo{pages}{1337--1350}.
	
	\bibitem[{May(2000)}]{may_role_2000}
	\bibinfo{author}{May, B.~J.} (\textbf{\bibinfo{year}{2000}}).
	\enquote{\bibinfo{title}{Role of the dorsal cochlear nucleus in the sound
			localization behavior of cats}} \bibinfo{journal}{Hear Res}
	\textbf{148}(1-2), \bibinfo{pages}{74--87}.
	
	\bibitem[{McAnally and Martin(2014)}]{mcanally_sound_2014}
	\bibinfo{author}{McAnally, K.~I.},  and \bibinfo{author}{Martin, R.~L.}
	(\textbf{\bibinfo{year}{2014}}).
	\enquote{\bibinfo{title}{\href{http://dx.doi.org/10.3389/fnins.2014.00210}{Sound
				localization with head movement: implications for 3-d audio displays}}}
	\bibinfo{journal}{Front Neurosci} \textbf{8}(210), \bibinfo{pages}{1--6}.
	
	\bibitem[{Meloni and Davis(1998)}]{meloni_dorsal_1998}
	\bibinfo{author}{Meloni, E.~G.},  and \bibinfo{author}{Davis, M.}
	(\textbf{\bibinfo{year}{1998}}). \enquote{\bibinfo{title}{The dorsal cochlear
			nucleus contributes to a high intensity component of the acoustic startle
			reflex in rats}} \bibinfo{journal}{Hear Res} \textbf{119}(1-2),
	\bibinfo{pages}{69--80}.
	
	\bibitem[{Mendon\c{c}a(2014)}]{mendonca_review_2014}
	\bibinfo{author}{Mendon\c{c}a, C.} (\textbf{\bibinfo{year}{2014}}).
	\enquote{\bibinfo{title}{\href{http://dx.doi.org/10.3389/fnins.2014.00219}{A
				review on auditory space adaptations to altered head-related cues}}}
	\bibinfo{journal}{Auditory Cogn Neurosci} \textbf{8}, \bibinfo{pages}{219}.
	
	\bibitem[{Mendon\c{c}a \emph{et~al.}(2013)Mendon\c{c}a, Campos, Dias, and
		Santos}]{mendonca_learning_2013}
	\bibinfo{author}{Mendon\c{c}a, C.}, \bibinfo{author}{Campos, G.},
	\bibinfo{author}{Dias, P.},  and \bibinfo{author}{Santos, J.~A.}
	(\textbf{\bibinfo{year}{2013}}).
	\enquote{\bibinfo{title}{\href{http://dx.doi.org/10.1371/journal.pone.0077900}{Learning
				{Auditory} {Space}: {Generalization} and {Long}-{Term} {Effects}}}}
	\bibinfo{journal}{PloS One} \textbf{8}(10), \bibinfo{pages}{e77900}.
	
	\bibitem[{Micheyl \emph{et~al.}(2007)Micheyl, Carlyon, Gutschalk, Melcher,
		Oxenham, Rauschecker, Tian, and Courtenay~Wilson}]{micheyl_role_2007}
	\bibinfo{author}{Micheyl, C.}, \bibinfo{author}{Carlyon, R.~P.},
	\bibinfo{author}{Gutschalk, A.}, \bibinfo{author}{Melcher, J.~R.},
	\bibinfo{author}{Oxenham, A.~J.}, \bibinfo{author}{Rauschecker, J.~P.},
	\bibinfo{author}{Tian, B.},  and \bibinfo{author}{Courtenay~Wilson, E.}
	(\textbf{\bibinfo{year}{2007}}).
	\enquote{\bibinfo{title}{\href{http://dx.doi.org/10.1016/j.heares.2007.01.007}{The
				role of auditory cortex in the formation of auditory streams}}}
	\bibinfo{journal}{Hear Res} \textbf{229}(1-2), \bibinfo{pages}{116--131}.
	
	\bibitem[{Middlebrooks(2009)}]{middlebrooks_auditory_2009}
	\bibinfo{author}{Middlebrooks, J.~C.} (\textbf{\bibinfo{year}{2009}}).
	\enquote{\bibinfo{title}{Auditory system: central pathways}} in
	\emph{\bibinfo{booktitle}{Encyclopedia of {Neuroscience}}}
	(\bibinfo{publisher}{Academic Press}, \bibinfo{address}{Oxford}), pp.
	\bibinfo{pages}{745--752}.
	
	\bibitem[{Middlebrooks(2015)}]{middlebrooks_sound_2015}
	\bibinfo{author}{Middlebrooks, J.~C.} (\textbf{\bibinfo{year}{2015}}).
	\enquote{\bibinfo{title}{\href{http://dx.doi.org/10.1016/B978-0-444-62630-1.00006-8}{Sound
				localization}}} \bibinfo{journal}{Handb Clin Neurol} \textbf{129},
	\bibinfo{pages}{99--116}.
	
	\bibitem[{Miller and Recanzone(2009)}]{miller_populations_2009}
	\bibinfo{author}{Miller, L.~M.},  and \bibinfo{author}{Recanzone, G.~H.}
	(\textbf{\bibinfo{year}{2009}}).
	\enquote{\bibinfo{title}{\href{http://dx.doi.org/10.1073/pnas.0901023106}{Populations
				of auditory cortical neurons can accurately encode acoustic space across
				stimulus intensity}}} \bibinfo{journal}{Proc Natl Acad Sci} \textbf{106}(14),
	\bibinfo{pages}{5931--5935}.
	
	\bibitem[{M{\o}ller \emph{et~al.}(1995)M{\o}ller, S{\o}rensen, Hammersh{\o}i,
		and Jensen}]{moller_head-related_1995}
	\bibinfo{author}{M{\o}ller, H.}, \bibinfo{author}{S{\o}rensen, M.~F.},
	\bibinfo{author}{Hammersh{\o}i, D.},  and \bibinfo{author}{Jensen, C.~B.}
	(\textbf{\bibinfo{year}{1995}}). \enquote{\bibinfo{title}{Head-related
			transfer functions of human subjects}} \bibinfo{journal}{J Audio Eng Soc}
	\textbf{43}, \bibinfo{pages}{300--321}.
	
	\bibitem[{M\"{o}ller and Raake(2014)}]{moller_quality_2014}
	\bibinfo{author}{M\"{o}ller, S.},  and \bibinfo{author}{Raake, A.}
	(\textbf{\bibinfo{year}{2014}}). \emph{\bibinfo{title}{Quality of
			{Experience}: {Advanced} {Concepts}, {Applications} and {Methods}}}
	(\bibinfo{publisher}{Springer}).
	
	\bibitem[{Moore \emph{et~al.}(2013)Moore, Brookes, and
		Naylor}]{moore_room_2013}
	\bibinfo{author}{Moore, A.~H.}, \bibinfo{author}{Brookes, M.},  and
	\bibinfo{author}{Naylor, P.~A.} (\textbf{\bibinfo{year}{2013}}).
	\enquote{\bibinfo{title}{Room geometry estimation from a single channel
			acoustic impulse response}} in \emph{\bibinfo{booktitle}{Proc Eur Signal
			Process Conf EUSIPCO}}, pp. \bibinfo{pages}{1--5}.
	
	\bibitem[{Morimoto \emph{et~al.}(2001)Morimoto, Iida, and
		Sakagami}]{morimoto_role_2001}
	\bibinfo{author}{Morimoto, M.}, \bibinfo{author}{Iida, K.},  and
	\bibinfo{author}{Sakagami, K.} (\textbf{\bibinfo{year}{2001}}).
	\enquote{\bibinfo{title}{\href{http://dx.doi.org/10.1016/S0003-682X(00)00051-7}{The
				role of reflections from behind the listener in spatial impression}}}
	\bibinfo{journal}{Appl Acoust} \textbf{62}(2), \bibinfo{pages}{109--124}.
	
	\bibitem[{Muniak and Ryugo(2014)}]{muniak_tonotopic_2014}
	\bibinfo{author}{Muniak, M.~A.},  and \bibinfo{author}{Ryugo, D.~K.}
	(\textbf{\bibinfo{year}{2014}}).
	\enquote{\bibinfo{title}{\href{http://dx.doi.org/10.1002/cne.23454}{Tonotopic
				organization of vertical cells in the dorsal cochlear nucleus of the
				{CBA}/{J} mouse: {Tonotopic} organization of vertical cells in the {DCN}}}}
	\bibinfo{journal}{J Comp Neurol} \textbf{522}(4), \bibinfo{pages}{937--949}.
	
	\bibitem[{Nelken \emph{et~al.}(2014)Nelken, Bizley, Shamma, and
		Wang}]{nelken_auditory_2014}
	\bibinfo{author}{Nelken, I.}, \bibinfo{author}{Bizley, J.},
	\bibinfo{author}{Shamma, S.~A.},  and \bibinfo{author}{Wang, X.}
	(\textbf{\bibinfo{year}{2014}}).
	\enquote{\bibinfo{title}{\href{http://dx.doi.org/10.1523/JNEUROSCI.2989-14.2014}{Auditory
				cortical processing in real-world listening: the auditory system going
				real}}} \bibinfo{journal}{J Neurosci} \textbf{34}(46),
	\bibinfo{pages}{15135--15138}.
	
	\bibitem[{Noble and Gatehouse(2006)}]{noble_effects_2006}
	\bibinfo{author}{Noble, W.},  and \bibinfo{author}{Gatehouse, S.}
	(\textbf{\bibinfo{year}{2006}}).
	\enquote{\bibinfo{title}{\href{http://dx.doi.org/10.1080/14992020500376933}{Effects
				of bilateral versus unilateral hearing aid fitting on abilities measured by
				the {Speech}, {Spatial}, and {Qualities} of {Hearing} scale ({SSQ})}}}
	\bibinfo{journal}{Int J Audiol} \textbf{45}(3), \bibinfo{pages}{172--181}.
	
	\bibitem[{Oberfeld and Kl\"{o}ckner-Nowotny(2016)}]{oberfeld_individual_2016}
	\bibinfo{author}{Oberfeld, D.},  and \bibinfo{author}{Kl\"{o}ckner-Nowotny, F.}
	(\textbf{\bibinfo{year}{2016}}).
	\enquote{\bibinfo{title}{\href{http://dx.doi.org/10.7554/eLife.16747}{Individual
				differences in selective attention predict speech identification at a
				cocktail party}}} \bibinfo{journal}{eLife} \textbf{5},
	\bibinfo{pages}{16747}.
	
	\bibitem[{Okano \emph{et~al.}(1998)Okano, Beranek, and
		Hidaka}]{okano_relations_1998}
	\bibinfo{author}{Okano, T.}, \bibinfo{author}{Beranek, L.~L.},  and
	\bibinfo{author}{Hidaka, T.} (\textbf{\bibinfo{year}{1998}}).
	\enquote{\bibinfo{title}{\href{http://dx.doi.org/10.1121/1.423955}{Relations
				among interaural cross-correlation coefficient ({IACCE}), lateral fraction
				({LFE}), and apparent source width ({ASW}) in concert halls}}}
	\bibinfo{journal}{J Acoust Soc Am} \textbf{104}(1),
	\bibinfo{pages}{255--265}.
	
	\bibitem[{P\"{a}tynen and Lokki(2016)}]{patynen_concert_2016}
	\bibinfo{author}{P\"{a}tynen, J.},  and \bibinfo{author}{Lokki, T.}
	(\textbf{\bibinfo{year}{2016}}).
	\enquote{\bibinfo{title}{\href{http://dx.doi.org/10.1121/1.4944038}{Concert
				halls with strong and lateral sound increase the emotional impact of
				orchestra music}}} \bibinfo{journal}{J Acoust Soc Am} \textbf{139}(3),
	\bibinfo{pages}{1214--1224}.
	
	\bibitem[{Peck(1996)}]{peck_visual-auditory_1996}
	\bibinfo{author}{Peck, C.~K.} (\textbf{\bibinfo{year}{1996}}).
	\enquote{\bibinfo{title}{Visual-auditory integration in cat superior
			colliculus: implications for neuronal control of the orienting response}}
	\bibinfo{journal}{Prog Brain Res} \textbf{112}, \bibinfo{pages}{167--177}.
	
	\bibitem[{Perrett and Noble(1995)}]{perrett_available_1995}
	\bibinfo{author}{Perrett, S.},  and \bibinfo{author}{Noble, W.}
	(\textbf{\bibinfo{year}{1995}}). \enquote{\bibinfo{title}{Available response
			choices affect localization of sound}} \bibinfo{journal}{Percept Psychophys}
	\textbf{57}, \bibinfo{pages}{150--158}.
	
	\bibitem[{Perrott and Saberi(1990)}]{perrott_minimum_1990}
	\bibinfo{author}{Perrott, D.~R.},  and \bibinfo{author}{Saberi, K.}
	(\textbf{\bibinfo{year}{1990}}).
	\enquote{\bibinfo{title}{\href{http://dx.doi.org/10.1121/1.399421}{Minimum
				audible angle thresholds for sources varying in both elevation and azimuth}}}
	\bibinfo{journal}{J Acoust Soc Am} \textbf{87}(4),
	\bibinfo{pages}{1728--1731}.
	
	\bibitem[{Rauschecker(2011)}]{rauschecker_expanded_2011}
	\bibinfo{author}{Rauschecker, J.~P.} (\textbf{\bibinfo{year}{2011}}).
	\enquote{\bibinfo{title}{\href{http://dx.doi.org/10.1016/j.heares.2010.09.001}{An
				expanded role for the dorsal auditory pathway in sensorimotor control and
				integration}}} \bibinfo{journal}{Hear Res} \textbf{271}(1-2),
	\bibinfo{pages}{16--25}.
	
	\bibitem[{Rauschecker and Tian(2000)}]{rauschecker_mechanisms_2000}
	\bibinfo{author}{Rauschecker, J.~P.},  and \bibinfo{author}{Tian, B.}
	(\textbf{\bibinfo{year}{2000}}).
	\enquote{\bibinfo{title}{\href{http://dx.doi.org/10.1073/pnas.97.22.11800}{Mechanisms
				and streams for processing of "what" and "where" in auditory cortex}}}
	\bibinfo{journal}{Proc Natl Acad Sci} \textbf{97}(22),
	\bibinfo{pages}{11800--11806}.
	
	\bibitem[{Reichinger \emph{et~al.}(2013)Reichinger, Majdak, Sablatnig, and
		Maierhofer}]{reichinger_evaluation_2013}
	\bibinfo{author}{Reichinger, A.}, \bibinfo{author}{Majdak, P.},
	\bibinfo{author}{Sablatnig, R.},  and \bibinfo{author}{Maierhofer, S.}
	(\textbf{\bibinfo{year}{2013}}).
	\enquote{\bibinfo{title}{\href{http://dx.doi.org/10.1109/3DV.2013.58}{Evaluation
				of {Methods} for {Optical} 3-{D} {Scanning} of {Human} {Pinnas}}}} in
	\emph{\bibinfo{booktitle}{Int {Conference} on 3D {Vision}}}, pp.
	\bibinfo{pages}{390--397}.
	
	\bibitem[{Ruggles \emph{et~al.}(2012)Ruggles, Bharadwaj, and
		Shinn-Cunningham}]{ruggles_why_2012}
	\bibinfo{author}{Ruggles, D.}, \bibinfo{author}{Bharadwaj, H.},  and
	\bibinfo{author}{Shinn-Cunningham, B.~G.} (\textbf{\bibinfo{year}{2012}}).
	\enquote{\bibinfo{title}{\href{http://dx.doi.org/10.1016/j.cub.2012.05.025}{Why
				{Middle}-{Aged} {Listeners} {Have} {Trouble} {Hearing} in {Everyday}
				{Settings}}}} \bibinfo{journal}{Curr Biol} \textbf{22}(15),
	\bibinfo{pages}{1417--1422}.
	
	\bibitem[{Salminen \emph{et~al.}(2015)Salminen, Takanen, Santala, Lamminsalo,
		Alto\`{e}, and Pulkki}]{salminen_integrated_2015}
	\bibinfo{author}{Salminen, N.~H.}, \bibinfo{author}{Takanen, M.},
	\bibinfo{author}{Santala, O.}, \bibinfo{author}{Lamminsalo, J.},
	\bibinfo{author}{Alto\`{e}, A.},  and \bibinfo{author}{Pulkki, V.}
	(\textbf{\bibinfo{year}{2015}}).
	\enquote{\bibinfo{title}{\href{http://dx.doi.org/10.1016/j.heares.2015.06.006}{Integrated
				processing of spatial cues in human auditory cortex}}} \bibinfo{journal}{Hear
		Res} \textbf{327}, \bibinfo{pages}{143--152}.
	
	\bibitem[{Schechtman \emph{et~al.}(2012)Schechtman, Shrem, and
		Deouell}]{schechtman_spatial_2012}
	\bibinfo{author}{Schechtman, E.}, \bibinfo{author}{Shrem, T.},  and
	\bibinfo{author}{Deouell, L.~Y.} (\textbf{\bibinfo{year}{2012}}).
	\enquote{\bibinfo{title}{\href{http://dx.doi.org/10.1523/JNEUROSCI.1315-12.2012}{Spatial
				{Localization} of {Auditory} {Stimuli} in {Human} {Auditory} {Cortex} is
				{Based} on {Both} {Head}-{Independent} and {Head}-{Centered} {Coordinate}
				{Systems}}}} \bibinfo{journal}{J Neurosci} \textbf{32}(39),
	\bibinfo{pages}{13501--13509}.
	
	\bibitem[{Schultz and Schultz(2015)}]{schultz_history_2015}
	\bibinfo{author}{Schultz, D.~P.},  and \bibinfo{author}{Schultz, S.~E.}
	(\textbf{\bibinfo{year}{2015}}). \emph{\bibinfo{title}{A {History} of
			{Modern} {Psychology}}}, \bibinfo{edition}{11} ed.
	(\bibinfo{publisher}{Cengage Learning}, \bibinfo{address}{Boston, MA}).
	
	\bibitem[{Shinn-Cunningham(2008)}]{shinn-cunningham_object-based_2008}
	\bibinfo{author}{Shinn-Cunningham, B.~G.} (\textbf{\bibinfo{year}{2008}}).
	\enquote{\bibinfo{title}{\href{http://dx.doi.org/10.1016/j.tics.2008.02.003}{Object-based
				auditory and visual attention}}} \bibinfo{journal}{Trends Cogn Sci}
	\textbf{12}(5), \bibinfo{pages}{182--186}.
	
	\bibitem[{Shinn-Cunningham \emph{et~al.}(2000)Shinn-Cunningham, Santarelli, and
		Kopco}]{shinn-cunningham_tori_2000}
	\bibinfo{author}{Shinn-Cunningham, B.~G.}, \bibinfo{author}{Santarelli, S.},
	and \bibinfo{author}{Kopco, N.} (\textbf{\bibinfo{year}{2000}}).
	\enquote{\bibinfo{title}{\href{http://view.ncbi.nlm.nih.gov/pubmed/10738816}{Tori
				of confusion: binaural localization cues for sources within reach of a
				listener}}} \bibinfo{journal}{J Acoust Soc Am} \textbf{107}(3),
	\bibinfo{pages}{1627--36}.
	
	\bibitem[{Shore(2005)}]{shore_multisensory_2005}
	\bibinfo{author}{Shore, S.~E.} (\textbf{\bibinfo{year}{2005}}).
	\enquote{\bibinfo{title}{\href{http://dx.doi.org/10.1111/j.1460-9568.2005.04142.x}{Multisensory
				integration in the dorsal cochlear nucleus: unit responses to acoustic and
				trigeminal ganglion stimulation}}} \bibinfo{journal}{Eur J Neurosci}
	\textbf{21}(12), \bibinfo{pages}{3334--3348}.
	
	\bibitem[{Singla \emph{et~al.}(2017)Singla, Dempsey, Warren, Enikolopov, and
		Sawtell}]{singla_cerebellum-like_2017}
	\bibinfo{author}{Singla, S.}, \bibinfo{author}{Dempsey, C.},
	\bibinfo{author}{Warren, R.}, \bibinfo{author}{Enikolopov, A.~G.},  and
	\bibinfo{author}{Sawtell, N.~B.} (\textbf{\bibinfo{year}{2017}}).
	\enquote{\bibinfo{title}{\href{http://dx.doi.org/10.1038/nn.4567}{A
				cerebellum-like circuit in the auditory system cancels responses to
				self-generated sounds}}} \bibinfo{journal}{Nat Neurosci} \textbf{20}(7),
	\bibinfo{pages}{943--950}.
	
	\bibitem[{Skottun \emph{et~al.}(2001)Skottun, Shackleton, Arnott, and
		Palmer}]{skottun_ability_2001}
	\bibinfo{author}{Skottun, B.~C.}, \bibinfo{author}{Shackleton, T.~M.},
	\bibinfo{author}{Arnott, R.~H.},  and \bibinfo{author}{Palmer, A.~R.}
	(\textbf{\bibinfo{year}{2001}}).
	\enquote{\bibinfo{title}{\href{http://dx.doi.org/10.1073/pnas.241513998}{The
				ability of inferior colliculus neurons to signal differences in interaural
				delay}}} \bibinfo{journal}{Proc Natl Acad Sci} \textbf{98}(24),
	\bibinfo{pages}{14050--14054}.
	
	\bibitem[{Slama and Delgutte(2015)}]{slama_neural_2015}
	\bibinfo{author}{Slama, M. C.~C.},  and \bibinfo{author}{Delgutte, B.}
	(\textbf{\bibinfo{year}{2015}}).
	\enquote{\bibinfo{title}{\href{http://dx.doi.org/10.1523/JNEUROSCI.3615-14.2015}{Neural
				{Coding} of {Sound} {Envelope} in {Reverberant} {Environments}}}}
	\bibinfo{journal}{J Neurosci} \textbf{35}(10), \bibinfo{pages}{4452--4468}.
	
	\bibitem[{Slater(2003)}]{slater_note_2003}
	\bibinfo{author}{Slater, M.} (\textbf{\bibinfo{year}{2003}}).
	\enquote{\bibinfo{title}{A {Note} on {Presence} {Terminology}}}
	\bibinfo{journal}{Presence Connect} \textbf{3}.
	
	\bibitem[{Slee and Young(2013)}]{slee_linear_2013}
	\bibinfo{author}{Slee, S.~J.},  and \bibinfo{author}{Young, E.~D.}
	(\textbf{\bibinfo{year}{2013}}).
	\enquote{\bibinfo{title}{\href{http://dx.doi.org/10.1523/JNEUROSCI.3437-12.2013}{Linear
				{Processing} of {Interaural} {Level} {Difference} {Underlies} {Spatial}
				{Tuning} in the {Nucleus} of the {Brachium} of the {Inferior} {Colliculus}}}}
	\bibinfo{journal}{J Neurosci} \textbf{33}(9), \bibinfo{pages}{3891--3904}.
	
	\bibitem[{Slee and Young(2014)}]{slee_alignment_2014}
	\bibinfo{author}{Slee, S.~J.},  and \bibinfo{author}{Young, E.~D.}
	(\textbf{\bibinfo{year}{2014}}).
	\enquote{\bibinfo{title}{\href{http://dx.doi.org/10.1152/jn.00885.2013}{Alignment
				of sound localization cues in the nucleus of the brachium of the inferior
				colliculus}}} \bibinfo{journal}{J Neurophysiol} \textbf{111}(12),
	\bibinfo{pages}{2624--2633}.
	
	\bibitem[{Snyder and Elhilali(2017)}]{snyder_recent_2017}
	\bibinfo{author}{Snyder, J.~S.},  and \bibinfo{author}{Elhilali, M.}
	(\textbf{\bibinfo{year}{2017}}).
	\enquote{\bibinfo{title}{\href{http://dx.doi.org/10.1111/nyas.13317}{Recent
				advances in exploring the neural underpinnings of auditory scene
				perception}}} \bibinfo{journal}{Ann NY Acad Sci} \textbf{1396}(1),
	\bibinfo{pages}{39--55}.
	
	\bibitem[{Sokolov(2001)}]{sokolov_orienting_2001}
	\bibinfo{author}{Sokolov, E.} (\textbf{\bibinfo{year}{2001}}).
	\enquote{\bibinfo{title}{\href{http://linkinghub.elsevier.com/retrieve/pii/B0080430767035361}{Orienting
				{Response}}}} in \emph{\bibinfo{booktitle}{International {Encyclopedia} of
			the {Social} \& {Behavioral} {Sciences}}} (\bibinfo{publisher}{Elsevier},
	\bibinfo{address}{Pergamon}), pp. \bibinfo{pages}{10978--10981}.
	
	\bibitem[{Strack and Deutsch(2004)}]{strack_reflective_2004}
	\bibinfo{author}{Strack, F.},  and \bibinfo{author}{Deutsch, R.}
	(\textbf{\bibinfo{year}{2004}}). \enquote{\bibinfo{title}{Reflective and
			{Impulsive} {Determinants} of {Social} {Behavior}}} \bibinfo{journal}{Pers
		Soc Psychol Rev} \textbf{8}(3), \bibinfo{pages}{220--247}.
	
	\bibitem[{Straka \emph{et~al.}(2014)Straka, Schmitz, and
		Lim}]{straka_response_2014}
	\bibinfo{author}{Straka, M.~M.}, \bibinfo{author}{Schmitz, S.},  and
	\bibinfo{author}{Lim, H.~H.} (\textbf{\bibinfo{year}{2014}}).
	\enquote{\bibinfo{title}{\href{http://dx.doi.org/10.1152/jn.00008.2014}{Response
				features across the auditory midbrain reveal an organization consistent with
				a dual lemniscal pathway}}} \bibinfo{journal}{J Neurophysiol}
	\textbf{112}(4), \bibinfo{pages}{981--998}.
	
	\bibitem[{Szab\'{o} \emph{et~al.}(2016)Szab\'{o}, Denham, and
		Winkler}]{szabo_computational_2016}
	\bibinfo{author}{Szab\'{o}, B.~T.}, \bibinfo{author}{Denham, S.~L.},  and
	\bibinfo{author}{Winkler, I.} (\textbf{\bibinfo{year}{2016}}).
	\enquote{\bibinfo{title}{\href{http://dx.doi.org/10.3389/fnins.2016.00524}{Computational
				{Models} of {Auditory} {Scene} {Analysis}: {A} {Review}}}}
	\bibinfo{journal}{Front Neurosci} \textbf{10}.
	
	\bibitem[{Trapeau \emph{et~al.}(2016)Trapeau, Aubrais, and
		Sch\"{o}nwiesner}]{trapeau_fast_2016}
	\bibinfo{author}{Trapeau, R.}, \bibinfo{author}{Aubrais, V.},  and
	\bibinfo{author}{Sch\"{o}nwiesner, M.} (\textbf{\bibinfo{year}{2016}}).
	\enquote{\bibinfo{title}{\href{http://dx.doi.org/10.1121/1.4960568}{Fast and
				persistent adaptation to new spectral cues for sound localization suggests a
				many-to-one mapping mechanism}}} \bibinfo{journal}{J Acoust Soc Am}
	\textbf{140}(2), \bibinfo{pages}{879--890}.
	
	\bibitem[{Trapeau and Sch\"{o}nwiesner(2015)}]{trapeau_adaptation_2015}
	\bibinfo{author}{Trapeau, R.},  and \bibinfo{author}{Sch\"{o}nwiesner, M.}
	(\textbf{\bibinfo{year}{2015}}).
	\enquote{\bibinfo{title}{\href{http://dx.doi.org/10.1016/j.neuroimage.2015.06.006}{Adaptation
				to shifted interaural time differences changes encoding of sound location in
				human auditory cortex}}} \bibinfo{journal}{NeuroImage} \textbf{118},
	\bibinfo{pages}{26--38}.
	
	\bibitem[{Viaud-Delmon and Warusfel(2014)}]{viaud-delmon_ear_2014}
	\bibinfo{author}{Viaud-Delmon, I.},  and \bibinfo{author}{Warusfel, O.}
	(\textbf{\bibinfo{year}{2014}}).
	\enquote{\bibinfo{title}{\href{http://dx.doi.org/10.3389/fnins.2014.00283}{From
				ear to body: the auditory-motor loop in spatial cognition}}}
	\bibinfo{journal}{Auditory Cogn Neurosci} \textbf{8}, \bibinfo{pages}{283}.
	
	\bibitem[{Vorl\"{a}nder and Shinn-Cunningham(2014)}]{hale_virtual_2014}
	\bibinfo{author}{Vorl\"{a}nder, M.},  and \bibinfo{author}{Shinn-Cunningham,
		B.} (\textbf{\bibinfo{year}{2014}}). \enquote{\bibinfo{title}{Virtual
			auditory displays}} in \emph{\bibinfo{booktitle}{Handbook of virtual
			environment technology}}, edited by \bibinfo{editor}{K.~S. Hale} and
	\bibinfo{editor}{K.~M. Stanney}, \bibinfo{edition}{2} ed.
	(\bibinfo{publisher}{CRC Press}, \bibinfo{address}{Boca Raton}), pp.
	\bibinfo{pages}{87--114}.
	
	\bibitem[{Whitmer \emph{et~al.}(2013)Whitmer, Seeber, and
		Akeroyd}]{whitmer_measuring_2013}
	\bibinfo{author}{Whitmer, W.~M.}, \bibinfo{author}{Seeber, B.~U.},  and
	\bibinfo{author}{Akeroyd, M.~A.} (\textbf{\bibinfo{year}{2013}}).
	\enquote{\bibinfo{title}{\href{http://dx.doi.org/10.1007/978-1-4614-1590-9_34}{Measuring
				the apparent width of auditory sources in normal and impaired hearing}}}
	\bibinfo{journal}{Advances in Experimental Medicine and Biology}
	\textbf{787}, \bibinfo{pages}{303--310}.
	
	\bibitem[{Winkler \emph{et~al.}(2009)Winkler, Denham, and
		Nelken}]{winkler_modeling_2009}
	\bibinfo{author}{Winkler, I.}, \bibinfo{author}{Denham, S.~L.},  and
	\bibinfo{author}{Nelken, I.} (\textbf{\bibinfo{year}{2009}}).
	\enquote{\bibinfo{title}{\href{http://dx.doi.org/10.1016/j.tics.2009.09.003}{Modeling
				the auditory scene: predictive regularity representations and perceptual
				objects}}} \bibinfo{journal}{Trends Cogn Sci} \textbf{13}(12),
	\bibinfo{pages}{532--540}.
	
	\bibitem[{Witmer and Singer(1998)}]{witmer_measuring_1998}
	\bibinfo{author}{Witmer, B.~G.},  and \bibinfo{author}{Singer, M.~J.}
	(\textbf{\bibinfo{year}{1998}}). \enquote{\bibinfo{title}{Measuring presence
			in virtual environments: {A} presence questionnaire}}
	\bibinfo{journal}{Presence} \textbf{7}(3), \bibinfo{pages}{225--240}.
	
	\bibitem[{Woods(1964)}]{woods_behavior_1964}
	\bibinfo{author}{Woods, J.~W.} (\textbf{\bibinfo{year}{1964}}).
	\enquote{\bibinfo{title}{\href{http://dx.doi.org/10.1152/jn.1964.27.4.635}{Behavior
				of chronic decerebrate rats}}} \bibinfo{journal}{J Neurophysiol} \textbf{27},
	\bibinfo{pages}{635--644}.
	
	\bibitem[{Xie(2013)}]{xie_head-related_2013}
	\bibinfo{author}{Xie, B.} (\textbf{\bibinfo{year}{2013}}).
	\emph{\bibinfo{title}{Head-related transfer function and virtual auditory
			display}} (\bibinfo{publisher}{J. Ross Publishing},
	\bibinfo{address}{Plantatation, FL}).
	
	\bibitem[{Yao \emph{et~al.}(2015)Yao, Bremen, and
		Middlebrooks}]{yao_transformation_2015}
	\bibinfo{author}{Yao, J.~D.}, \bibinfo{author}{Bremen, P.},  and
	\bibinfo{author}{Middlebrooks, J.~C.} (\textbf{\bibinfo{year}{2015}}).
	\enquote{\bibinfo{title}{\href{http://dx.doi.org/10.1152/jn.01029.2014}{Transformation
				of spatial sensitivity along the ascending auditory pathway}}}
	\bibinfo{journal}{J Neurophysiol} \textbf{113}(9),
	\bibinfo{pages}{3098--3111}.
	
	\bibitem[{Yost(1974)}]{yost_discriminations_1974}
	\bibinfo{author}{Yost, W.~A.} (\textbf{\bibinfo{year}{1974}}).
	\enquote{\bibinfo{title}{\href{http://dx.doi.org/10.1121/1.1914701}{Discriminations
				of interaural phase differences}}} \bibinfo{journal}{J Acoust Soc Am}
	\textbf{55}(6), \bibinfo{pages}{1299--1303}.
	
	\bibitem[{Yost \emph{et~al.}(2015)Yost, Zhong, and Najam}]{yost_judging_2015}
	\bibinfo{author}{Yost, W.~A.}, \bibinfo{author}{Zhong, X.},  and
	\bibinfo{author}{Najam, A.} (\textbf{\bibinfo{year}{2015}}).
	\enquote{\bibinfo{title}{\href{http://dx.doi.org/10.1121/1.4935091}{Judging
				sound rotation when listeners and sounds rotate: {Sound} source localization
				is a multisystem process}}} \bibinfo{journal}{J Acoust Soc Am}
	\textbf{138}(5), \bibinfo{pages}{3293--3310}.
	
	\bibitem[{Ziegelwanger \emph{et~al.}(2016)Ziegelwanger, Kreuzer, and
		Majdak}]{ziegelwanger_priori_2016}
	\bibinfo{author}{Ziegelwanger, H.}, \bibinfo{author}{Kreuzer, W.},  and
	\bibinfo{author}{Majdak, P.} (\textbf{\bibinfo{year}{2016}}).
	\enquote{\bibinfo{title}{\href{http://dx.doi.org/10.1016/j.apacoust.2016.07.005}{A
				priori mesh grading for the numerical calculation of the head-related
				transfer functions}}} \bibinfo{journal}{Appl Acoust} \textbf{114},
	\bibinfo{pages}{99--110}.
	
	\bibitem[{Ziegelwanger and Majdak(2014)}]{ziegelwanger_modeling_2014}
	\bibinfo{author}{Ziegelwanger, H.},  and \bibinfo{author}{Majdak, P.}
	(\textbf{\bibinfo{year}{2014}}).
	\enquote{\bibinfo{title}{\href{http://dx.doi.org/10.1121/1.4863196}{Modeling
				the direction-continuous time-of-arrival in head-related transfer
				functions}}} \bibinfo{journal}{J Acoust Soc Am} \textbf{135}(3),
	\bibinfo{pages}{1278--1293}.
	
	\bibitem[{Ziegelwanger \emph{et~al.}(2015)Ziegelwanger, Majdak, and
		Kreuzer}]{ziegelwanger_numerical_2015}
	\bibinfo{author}{Ziegelwanger, H.}, \bibinfo{author}{Majdak, P.},  and
	\bibinfo{author}{Kreuzer, W.} (\textbf{\bibinfo{year}{2015}}).
	\enquote{\bibinfo{title}{\href{http://dx.doi.org/10.1121/1.4922518}{Numerical
				calculation of listener-specific head-related transfer functions and sound
				localization: {Microphone} model and mesh discretization}}}
	\bibinfo{journal}{J Acoust Soc Am} \textbf{138}(1),
	\bibinfo{pages}{208--222}.
	
\end{thebibliography}
\end{document}